\documentclass[11pt]{article}

\usepackage{amsmath, amssymb, amsfonts, amsthm}
\usepackage[dvipsnames]{xcolor}
\usepackage{graphicx}
\usepackage[]{subfigure}
\usepackage[font=footnotesize,format=plain,justification=justified,width=\textwidth]{caption}
\usepackage{tikz}
\usepackage{pgfplots}
\usepackage{pgfplotstable}
\usepgfplotslibrary{patchplots}
\usepgfplotslibrary{groupplots}
\usepgfplotslibrary{fillbetween}

\usepackage{scalefnt}
\usepackage{microtype}
\usepackage{xfrac}
\usepackage{relsize}

\pgfplotsset{compat=1.9}

\pgfkeys{/pgf/number format/.cd,1000 sep={\,}}
\pgfplotsset{
  every tick label/.append style={scale=0.8}
}%


\newcount\ndots
\def\drawline#1#2{\raise 2.5pt\vbox{\hrule width #1pt height #2pt}}

\def\trian{\raise 1.25pt\hbox{$\scriptscriptstyle\triangle$}\nobreak\ }
\def\solidtrian{\raise 1.25pt
\hbox to 3bp{
\hfill}\nobreak\ }
\def\dsolidtrian{\raise 1.25pt
\hbox to 3bp{
\hfill}\nobreak\ }
\def\soliddiamond{\raise 1.25pt
\hbox to 4bp{
\hfill}\nobreak\ }

\def\square{${\vcenter{\hrule height .4pt
              \hbox{\vrule width .4pt height 3pt \kern 3pt \vrule width .4pt}
          \hrule height .4pt}}$\nobreak\ }

\def\plus{\raise 1.25pt \hbox{$\scriptscriptstyle +$}\nobreak\ }
\def\x{\raise 1.25pt \hbox{$\scriptscriptstyle \times$}\nobreak\ }
\def\legendtable#1{\vbox{\baselineskip=10pt\tabskip=0pt\let\\=\cr\halign{\hfil##\hskip 3pt&##\hfil\cr#1\crcr}}}
\def\lllegend#1 #2 #3{\figlab {#1} {#2} {\legendtable{#3}}}
\def\lrlegend#1 #2 #3{\figlab {#1} {#2} {\llap{\legendtable{#3}}}}
\def\ullegend#1 #2 #3{\figlab {#1} {#2} {\vtop{\hrule height 0pt\legendtable{#3}}}}
\def\urlegend#1 #2 #3{\figlab {#1} {#2} {\llap{\vtop{\hrule height
0pt\legendtable{#3}}}}}

\def\Dpartial#1#2{ \frac{\partial #1}{\partial #2} }

\DeclareMathOperator{\sgn}{sgn}

\def\onedot{$\mathsurround0pt\ldotp$}
\def\cddot{%
  \mathbin{\vcenter{\baselineskip.67ex%
    \hbox{\onedot}\hbox{\onedot}}%
  }}%

\newcommand{\bD}{{\mathbf{D}}}
\newcommand{\bE}{{\mathbf{E}}}
\newcommand{\bF}{{\mathbf{F}}}
\newcommand{\bI}{{\mathbf{I}}}
\newcommand{\bP}{{\mathbf{P}}}

\newcommand{\bS}{{\mathbf{S}}}
\newcommand{\bU}{{\mathbf{U}}}

\newcommand{\bsigma}{{\boldsymbol{\sigma}}}

\newcommand{\be}{{\mathbf{e}}}
\newcommand{\bn}{{\mathbf{n}}}
\newcommand{\bbf}{{\mathbf{f}}}
\newcommand{\bp}{{\mathbf{p}}}
\newcommand{\br}{{\mathbf{r}}}
\newcommand{\bu}{{\mathbf{u}}}

\newcommand{\bx}{{\mathbf{x}}}

\newcommand{\shalf}{\smash{\frac{1}{2}}}
\newcommand{\ero}{{\mathbf{e}_{r_o}}}
\newcommand{\req}{{r_o^{\text{eq}}}}

\definecolor{RYB1}{RGB}{207, 37, 37}
\definecolor{RYB2}{RGB}{37, 91, 207}
\definecolor{RYB3}{RGB}{37, 207, 91}
\definecolor{RYB4}{RGB}{163,26,145}
\definecolor{RYB5}{RGB}{253, 180, 98}
\definecolor{RYB6}{RGB}{179, 222, 105}
\definecolor{RYB7}{RGB}{128, 177, 211}

\pgfplotscreateplotcyclelist{newcolors}{
{RYB1,every mark/.append style={fill=RYB1,mark size={2.5}},mark=*},
{RYB2,every mark/.append style={fill=RYB2},mark=square*},
{RYB3,every mark/.append style={fill=RYB3,mark size={3}},mark=triangle*},
{RYB4,every mark/.append style={fill=RYB4,mark size={4}},mark=x},
{RYB5,every mark/.append style={fill=RYB5,mark size={3}},mark=oplus},
{RYB7,every mark/.append style={fill=RYB7},mark=*},
}

\begin{document}

\title{\Large\textbf{ A free object in a confined active contractile nematic fluid:  fixed-point and limit-cycle behaviors}}

\author{Jonathan B.~Freund\footnote{Department of Aerospace Engineering,
  University of Illinois Urbana--Champaign, jbfreund@illinois.edu}}

\date{\today}

\maketitle

\begin{abstract}
Simulations employing the continuum model of Gao \textit{et al.}\ [\textit{Phy.\ Rev.\ Fluids}, \textbf{2} 093302 (2017)] are used to study the transport of an object in a closed two-dimensional container by a dense suspension of contractile active agents.  For parameters that generally yield nematic alignment, the initial flow and object motion is typically characterized by chaotically interacting $m=\pm\shalf$ defects in its nematic structure, which form in oppositely signed pairs or on the container wall or on the object.  Those that form on the object also make an oppositely-signed contribution to the $m_\circ$ nematic structure associated with the object.  However, in many cases, the chaotic flow does not persist.  It instead ends up in one of two states, which are studied in detail for a circular object in a circular container.  One is a fixed point, associated with a $m_\circ=+1$ object with radial nematic ordering.  The suspension flows but with the object stationary near the container wall.  A sharply-aligned nematic model confirms that its position is maintained by a (nearly) hydrostatic balance and that a related circumferential $m_\circ=+1$ configuration, which is not observed, would indeed be unstable.  The second terminal behavior, which can occur for the same physical parameters as the fixed-point behavior, is associated with a $m_\circ=0$ object.  It is a limit-cycle oscillation in which the object cyclically traverses the container, spawning transient $m=-\shalf$ defect pairs each half-cycle.  Both of these configurations are analyzed in detail, and are potential related to simple biological tasks.  It is shown that they also occur in square and elliptical containers, with the ellipse displaying a particularly rich phenomenology that includes switching between them.
\end{abstract}

\section{Introduction} \label{s:intro}

Living systems are out of equilibrium but typically maintain a structure, or at least some manner of organization, and conduct tasks.  Biology and synthetic biology motivate investigation of how suspensions of microscopic agents---often macromolecular structures linked by motor proteins or mobile cells---give rise to higher-level behaviors, such as organized motion in circulating flow \cite{Dombrowski:2004,Hardouin:2020}, mixing in low-Reynolds-number flow \cite{Tan:2019}, the driven motion of objects \cite{Foffano:2012,Freund:2023}, and driven boundaries \cite{Furthauer:2012,Ray:2023}.  In this work, we are primarily interested in free objects and the mechanism and outcomes of their transport by an active suspension.

Mathematical descriptions of active fluid suspensions are being developed based on their active agent microstructure via multiple strategies \cite{Marchetti:2013}.  We focus on low-Reynolds-number hydrodynamically interacting active agents that are smaller than other scales in the system, as appropriate for many biological materials.  Models that have been developed for such systems are based on hydrodynamically and stericly interacting, strain-rate responsive, active dipoles, as models for swimmers or rigid macromolecules linked with motor proteins \cite{Saintillan:2008,Ezhilan:2013}.  Low-degree moments of a Smoulokowski equation description of the micro-kinetics yields a continuum model with an advected orientation tensor $\bD$ as the principal dependent variable and facilitates simulations in complex geometries \cite{Gao:2017,Weady:2022}.  Extensive analysis has shown that the averaging procedures leading to the tensor continuum model preserves key features of the more complete kinetic description, including its instabilities.

With these models now established in a self-consistent and phenomenologically realistic form, we focus on how objects are moved by such a suspension, and particularly if and how they achieve an ultimate state.  The action of nematic active suspensions on free particles \cite{Loewe:2022} or rotationally free objects \cite{Thampi:2016b,Ray:2023} generally show rich behaviors.  Particles that induce $m=+\shalf$ topological nematic structures on their surface can bind a nearby $-\shalf$ defect in a way that sustains particle motion \cite{Loewe:2022}.  Arrays of rotationally unconstrained disks arrive at a state of alternating spins that could extract energy from the fluid \cite{Thampi:2016b}, as can disks decorated with chiral features \cite{Ray:2023}.  Similarly, driven boundaries can induce defect dynamics and local organization of an active nematic material \cite{Rivas:2020}. Although the dynamics for most of the cases we consider are chaotic, we are primarily interested in ultimate outcomes, corresponding loosely with biological tasks.  In particular, we consider how the stochastic microscopic agents can collectively move the object to some location or in a particular manner.  The completion of such tasks is, in a sense, a foundation of life, so it is hoped that task-like outcomes might illuminate biological processes in systems similar to those that inspire the active suspension models we consider.

To move an object from an initial position, it is essential that the initial configuration be, in some sense, unstable, not a stable fixed-point solution of the dynamics. It is well-understood that instabilities of these and similar active suspensions are long wave-length \cite{Marchetti:2013,Ezhilan:2013}, which is why suspensions that spontaneously flow chaotically in large containers produce organized circulations in smaller containers and stabilize in still smaller containers \cite{Theillard:2017,Gao:2017}.   In a sufficiently small container with an immersed object within it, the mobility of the object can be essential for instability of the suspension, so that when mobility of the object is viscously restricted as it approaches the container wall, the suspension can stabilize.  That is, without any trigger, the suspension can switch from a state of transporting the object to leaving it in a fixed position near a container wall.  This was a key finding of a previous study, leading to the present effort \cite{Freund:2023}.  In that case, which neglected steric interactions, sufficient active extender strength yields instability, so this behavior depended on extensors (non-motile pushers) of sufficient strength, but not so strong as for them to be unstable for an immobile object.  

\begin{figure}
  \begin{center}
    \includegraphics[width=0.4\textwidth]{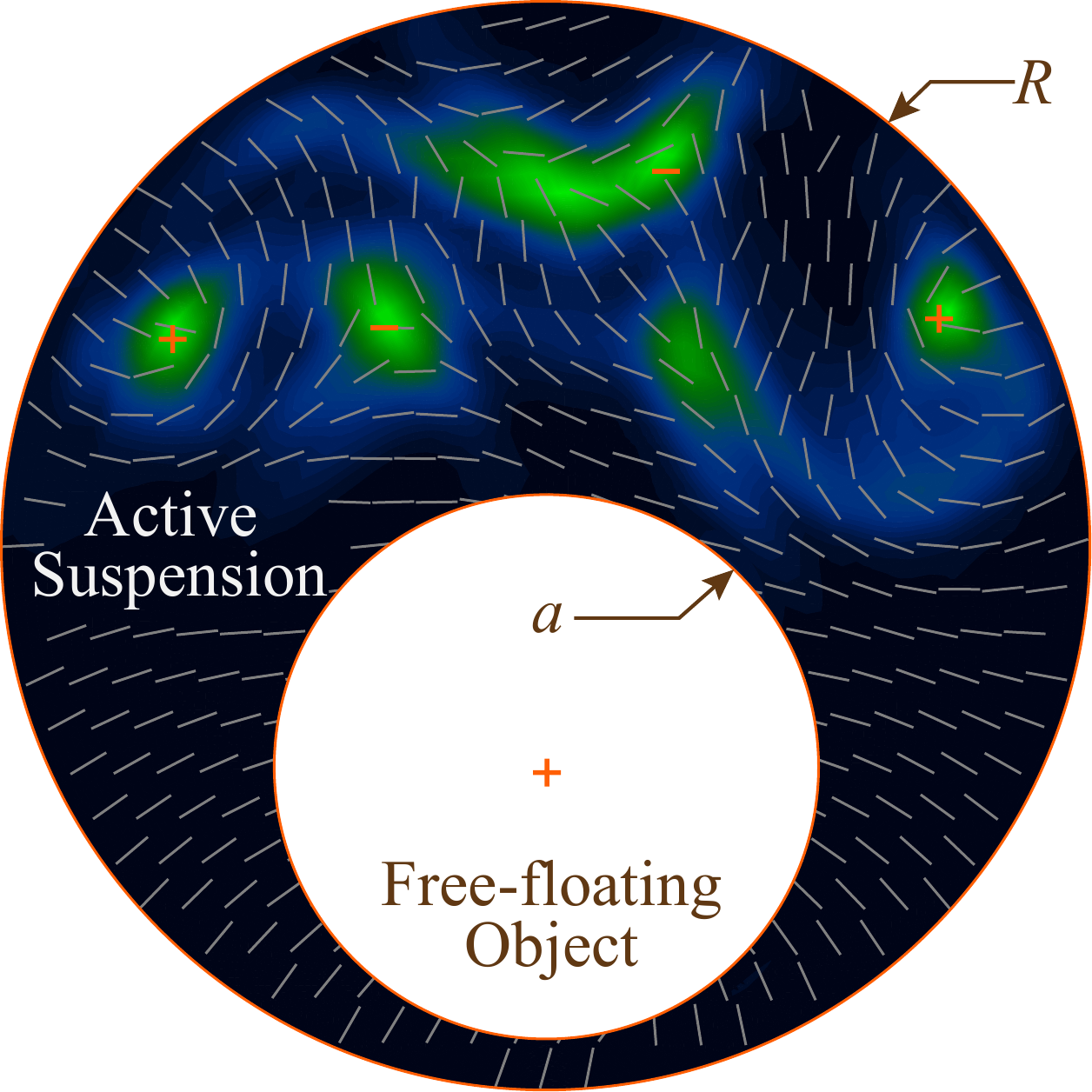}
  \end{center}
  \caption{Configuration schematic with visualized simulation result for $\zeta = 1$, $\alpha = 0.7$.  The $+/-$ symbols indicate $\pm \shalf$ defects in the nematic structure, and the $+$ on the object indicates its nematic charge $m_\circ=+\shalf$.  The colors indicate local degree of alignment, with black indicating $\lambda_1 =0$ and green indicating $\lambda_1 = 0.5$, where $\lambda_1$ is the smaller eigenvalue of the director tensor $\bD$.  The orientation of the gray segments indicate the local alignment direction $\theta$ at their midpoint. }\label{fig:schematic}
\end{figure}

For stericly interacting active agents, the suspension stability is more complex, and there is also potential coupling with the geometry via any nematic structure of the fluid.  As for diffuse suspensions, interacting extender agents are unstable, but with steric interactions, contractors can also be unstable \cite{Ezhilan:2013}.  Simplistically, the contractors need to be sufficiently strong to overcome viscous resistance and the nematic structure induced by their steric interactions.  Ezhilan \textit{et al.}\ \cite{Ezhilan:2013} provide a comprehensive stability analysis of their kinetic model, whose key features are well-represented by the continuum model we employ \cite{Weady:2022}.  There are ranges of parameters for which  a uniform base flow is unstable \cite{Freund:2023}. More complex cases are explored here in which an initially chaotic flow can lead to stable finite-amplitude fixed-point and limit-cycle behaviors.

Dense interacting suspensions of apolar rod-like particles share characteristics with nematic liquid crystals \cite{Giomi:2013}. Stacking defects (disclinations) in such systems arise as half-order ($m \pm \shalf$) charge-like singularities.  In inactive systems, out-of-equilibrium configurations are mediated by the boundaries and initial conditions, and defects are central to the overall stresses in the material.  The forces on an immersed object in such a fluid depends in complex ways on its shape and the locations and character of any defects \cite{Chandler:2023}.  Activity of the constituent agents complicates this with potential for driven formation, transport, and annihilation of defects \cite{Thampi:2016,Giomi:2013}.  For strong nematic alignment, defect charge is conserved, so they form and annihilate in oppositely signed pairs or at wall boundaries.  When they annihilate on an immersed object, such as we consider, the object's net nematic ordering $m_\circ$ also changes to conserve overall charge.   The two main solution classes found can arise upon the object attaining $m_\circ = +1$ versus $m_\circ=0$ nematic structure.

The following section~\ref{s:simmod} introduces the simulation model, including the configuration geometry, specifics of the active nematic fluid, the physical parameters and their associated time scales, and the numerical methods used.  The character of the chaotic flow that results from typical initial conditions, leading to the ultimate stable solutions, is briefly described in section~\ref{s:initialtraj}.  It involves the formation and annihilation of $m=\pm \shalf$ defects.  The ultimate stable fixed points and limit cycles are then analyzed in detail in sections~\ref{s:fixed-point} and \ref{s:limit-cycle}, respectively.  These include descriptions of the configurations, how they arise, and investigations of the ranges of physical parameters for which they occur.  For clarity, both of these sections focus on a circular object in a circular container, but similar behavior also arises in more complex containers, as demonstrated with examples in section~\ref{s:other}.  Implications of these results and directions forward from this study are revisited and discussed in section~\ref{s:conclusions}. 

\section{Simulation Model}\label{s:simmod}

\subsection{Configuration}

The configuration for most cases considered is shown in figure~\ref{fig:schematic}:  it is two dimensional, with a free-floating neutrally buoyant circular object (radius $a=1$) immersed in a circular container (radius $R=2$).  An active suspension fills the space between the container and the object.  Such a two-dimensional active suspension can be formed on an oil-water interface to which a microtubule-kinesin solution becomes bound, producing a quasi-two-dimensional flow \cite{Sanchez:2012,Rivas:2020,Ray:2023}.  This two-dimensional system is also a resonable model for identifying and understanding potential phenomena to  inform understanding of more complex analogues.  Reynolds numbers are assumed to be sufficiently small that inertia can be neglected.  Taking our two-dimensional simulation as an analoge of quasi-two-dimensional interface-bound suspensions is consistent with the in-plane interfacial stresses dominating viscous resistance in the fluids on either side of the interface.  In a few cases, the typically free object is instead moved at a constant velocity or held fixed in order to analyze mechanisms.

\subsection{Governing equations}
\label{s:formulation} 

The active suspension is assumed to have densely and uniformly distributed rod-like nematic contractors (immotile pullers) whose steric interactions are modeled with the Maier--Saupe potential \cite{Maier:1958}.  It is well-understood that Doi--Edwards \cite{Doi:1988} moments of a distribution function $\Psi(\bx,\bp,t)$ describing the local agent orientation $\bp$ in space $\bx$ and time $t$ for such system yields a continuum model for a director tensor field
\begin{equation}
  \bD(\bx,t) =  \int_{|\bp|=1} \bp\bp\, \Psi(\bx,\bp,t) \, d\bp
\end{equation}
that includes an unclosed fourth-moment rank-four tensor
\begin{equation}
  \bS(\bx,t) = \int_{|\bp|=1} \bp\bp \bp\bp\, \Psi(\bx,\bp,t) \, d\bp.
\end{equation}
Following exactly the development of Gao \textit{et al.}~\cite{Gao:2017}, which was subsequently used by Young \textit{et al.}~\cite{Young:2021} and is a special case of the more recent general continuum model for apolar non-uniformly distributed active systems of Weady \textit{et al.}~\cite{Weady:2022}, the momentum balance is
\begin{equation}
  -2 \nabla\cdot \bE + \nabla p = \nabla \cdot \big[\underbrace{\alpha \bD + \beta \bS \cddot \bE - 2 \beta \zeta (\bD\cdot\bD - \bS \cddot \bD)}_{\bsigma_a}\big]\label{e:mom}                   
\end{equation}
with the incompressiblity constraint
\begin{equation}
  \nabla \cdot \bu = 0 \label{e:incomp},                    
\end{equation}
and the $\bD$ field is governed by 
\begin{equation}
  \Dpartial{\bD}{t} + \bu \cdot \nabla \bD - (\nabla\bu \cdot \bD +
  \bD \cdot \nabla \bu^T) +  2 \bS \cddot \bE=   4\zeta (\bD\cdot\bD -
  \bS \cddot \bD) +  d_T \nabla^2
  \bD - 4 d_R(\bD - \bI/2),
  \label{e:D}
\end{equation}
where $\bI$ is identity (for isotropy), and $\bE$ is the strain-rate tensor:
\begin{equation}
\bE = \frac{1}{2}\left(\nabla \bu + \nabla \bu^T\right).
\end{equation}
Boundaries are no slip with $\bu = \bu_{\text{wall}}$ and induce no structural direction, so $\bn\cdot\nabla \bD = 0$, where $\bn$ is the wall unit normal.  The parameters appearing in (\ref{e:mom}) and (\ref{e:D}) are discussed in the following section.

To close (\ref{e:D}) with a model for $\bS$, Gao \textit{et al.}~\cite{Gao:2017}\ introduce a Bingham statistical model \cite{Chaubal:1998,Bingham:1974}, which importantly can represent both the isotropic and full-aligned nematic limits.  It also linearizes to be consistent with the underlying kinetic theory.  We use this same closure.  Weady \textit{et al.}~\cite{Weady:2022} generalize it to polar systems (the $B$-closure) and provide additional discussion about it.  They also introduce a fast algorithm for its numerical evaluation \cite{Weady:2022a}, though in two dimensions it seems prudent to employ an implicit analytic form cast in terms of modified Bessel functions \cite{Freund:2023}. Given iteration seeds from the previous time step, the resulting formula can be solved very accurately with few Newton iterations.  Near a condition of full alignment, where convergence of the Newton method slows, it can be evaluated explicitly in an accurate closed-form second-order asymptotic limit.

The free-floating condition corresponds to having zero net force and torque on the object,
\begin{align}
  \bF &= \int_{\circ} \big[-p \bI + (\nabla \bu + \nabla \bu^T) + \bsigma_a\big]\cdot \bn\,d\bx = \mathbf{0} \label{e:F}\\
  T &=\int_{\circ} \br \times\big( \big[-p \bI + (\nabla \bu + \nabla \bu^T) + \bsigma_a \big]\cdot \bn\big)\,d\bx =0 \label{e:T},
\end{align}
though in some cases the object's velocity is instead specified.

\subsection{Parameters and regime}

As discussed by Gao~\textit{et al.}~\cite{Gao:2017}, the non-dimensional parameters in (\ref{e:mom}) and (\ref{e:D}) have well-understood connections with microscopic mechanics of the solution.  Here we only summarize their meaning in their non-dimensional forms.  The strength of the activity is parameterized by $\alpha$; in a biological suspension, this would typically be related to the consumption of ATP.  Our focus on contractor agents, which introduce a tension along their axis, indicate $\alpha > 0$.  We take $\alpha \in [0,10]$, with a typical value of about $5$.  The steric interaction between the rod-like particles is parameterized by $\zeta$, with larger $\zeta$ promoting alignment.  We consider $\zeta \in [0,8]$, with $0.7$ and $1.0$ as typical values.  The effects we consider depend on sustained steric alignment, which requires that $\zeta$ be stronger than diffusion toward isotropy, which is parameterized by diffusivity $d_R=0.02$ in (\ref{e:D}).  %
The other diffusion term in (\ref{e:D}) acts on spatial $\bD$ gradients, and is taken to have a coefficient $d_T \approx 0.01$ in most cases.  The final parameter is $\beta$ in (\ref{e:mom}), which quantifies how the strain and steric interactions result in internal fluid stresses.  In all cases, we take $\beta = 0.5$.

Specific parameter values will be introduced as cases are presented, though we can anticipate in advance the overall regimes of behavior based on the mechanism time scales associated with the parameter values considered.  To make this assessment for our configuration we introduce two length scales.  The first the large spatial scale of the fluid $\ell$.  For an $R = 2$ container with a $a=1$ object, $\ell = 1$.  We anticipate a second relevant length scale to arise when the fluid becomes confined as the object approaches the container wall.  We take this to simply be $\delta = 0.1a = 0.1$, which the simulations will confirm to be relevant.  The key aspect of $\delta$ is that it is significantly smaller than the container scale.  We also anticipate that $\delta$ will be a defect-neighborhood scale.

We also introduce two velocity scales for defining time scales associated with the left-hand-side terms in (\ref{e:D}).  It might be possible to set the physical parameters so that velocities should arise such that certain mechanisms come into balance.  However, it is more straightforward (and a better reflection of how this study was developed) to appeal in advance to the forthcoming results, which will show that $u_\circ=0.01$ is a typical object velocity and $u_f = 0.1$ is a typical suspension velocity.  

Table~\ref{tab:taus} summarizes time scales corresponding to the mechanisms represented in (\ref{e:D}).  The nematic ordering carries the fastest time scale $\tau_\zeta$.   Rotational diffusion time being significantly slower than the nematic ordering stabilizes the nematic structure:  $\tau_R/\tau_\zeta  = 50\; (=\xi / 2)$, where $\xi > 8$ is sufficient for stable uniform nematic structure \cite{Gao:2015,Gao:2017}.  Another anticipated behavior is that thermal diffusion will be unimportant with respect to nematic ordering on the $\ell$-scale of the container, with $\tau_T^\ell/\tau_\zeta = 100$, so gradients in the nematic ordering should be able to persist across the container.  In contrast, thermal diffusion will be a significant factor in a narrow gap or any $\delta$-scale feature, with $\tau_T^\delta/\tau_\zeta = 1$.  This is consistent with a $\delta$-scale regularization of defect neighborhoods.  For the selected velocity scales, we anticipate diffusion to be able to couple with the container-scale object advection dynamics, at least under some circumstances, with $\tau_T^\ell/\tau_{u_\circ}^\ell = 1$.  The modest mismatch of flow-driven advective times with respect to nematic ordering and $\delta$-scale diffusion point to potential weak coupling in large regions, with  $\tau_{u_f}^\ell/\tau_\zeta = \tau_{u_f}^\ell/\tau_T^\delta = 10$.  However, these are expected to couple strongly in small (presumably high-strain-rate) regions, with $\tau_{u_f}^\delta/\tau_\zeta = \tau_{u_f}^\delta/\tau_T^\delta = 1$.

\begin{table}
  \begin{center}
  \begin{tabular}{rccc}\hline\\[-3mm]
    Container-scale object advection  & $\tau_{u_\circ}^\ell \equiv \frac{\ell}{u_\circ}$& $ \approx$& $100$ \\[2mm]
    Container-scale advection  & $\tau_{u_f}^\ell \equiv \frac{\ell}{u_f}$& $ \approx$& $10$ \\[2mm]
    Gap-/defect-scale advection  & $\tau_{u_f}^\delta \equiv \frac{\delta}{u_f}$& $ \approx$& $1$ \\[2mm]
    Nematic ordering  & $\tau_{\zeta} \equiv \frac{1}{\zeta}$& $ \approx$& $1$ \\[2mm]
    Container-scale diffusion  & $\tau_{T}^\ell \equiv \frac{\ell^2}{d_T}$& $ \approx$ & $100$ \\[2mm]
    Gap-/defect-scale diffusion  & $\tau_{T}^\delta \equiv \frac{\delta^2}{d_T}$& $ \approx$ & $1$ \\[2mm]
    Rotational diffusion  & $\tau_{R} \equiv \frac{1}{d_R}$& $=$ & $50$ \\[2mm]
    \hline
  \end{tabular}
  \caption{Mechanism time scales for nominal suspension parameter values and the observed velocities.  The non-dimensionalization and non-dimensional parameters follow directly from previous efforts \cite{Gao:2017, Young:2021}; see the text for discussion of time scales in the current context.}\label{tab:taus}
  \end{center}
\end{table}

\subsection{Numerical discretization}

Following a previous effort \cite{Freund:2023}, third-order continuous Galerkin finite elements are used to represent the $\bu$ and $\bD$ fields in (\ref{e:mom}) and (\ref{e:D}) on a mesh that advects with the local fluid velocity.  The pressure is treated as a Lagrange multiplier enforcing incompressibility (\ref{e:incomp}), and it is discretized as usual with a one degree lower basis, second-order elements in this case.  The advecting mesh incorporates $\bu\cdot\nabla\bD$ into the time derivative and facilitates tracking of the object motion.  Most cases used 4880 total degrees of freedom for $\bu$, $p$, and the two independent components of $\bD$.  An additional 3120 degrees of freedom represent the third-order mapping function describing the advected mesh.   This might seem like a small mesh, but convergence was confirmed to be rapid, as expected for spectral elements discretizing the smooth fields that arise for the parameters considered.  (Significantly more resolution was required in a previous effort, where more details about the numerical methods are also provided \cite{Freund:2023}.)  Time integration of (\ref{e:D}) is via a second-order explicit backward differencing scheme for all but the $d_T$ diffusion term, which is solved implicitly for the next time level in conjunction with inverting the so-called mass-matrix arising from the weak-form of the time derivative term.  Every ten numerical time steps, $\bD$ is projected onto a new mesh constructed of straight lines extending from the object to the container wall and a family of circles filling the region between the object and the container.  (The more complex geometries introduced in section~\ref{s:other} employ a straightforward generalization of this approach.)  The solver with this resolution was confirmed to reproduce the complex rotating flow in a circle found by Gao~\textit{et al.}, which is visualized in their figure number 10 \cite{Gao:2017}.   For the object motion, the linearity of (\ref{e:mom}) is used to solve for fields with zero object net force (\ref{e:F}) and torque (\ref{e:T}) as discretized with the same finite-element description.

\section{Chaotic Evolution}
\label{s:initialtraj}

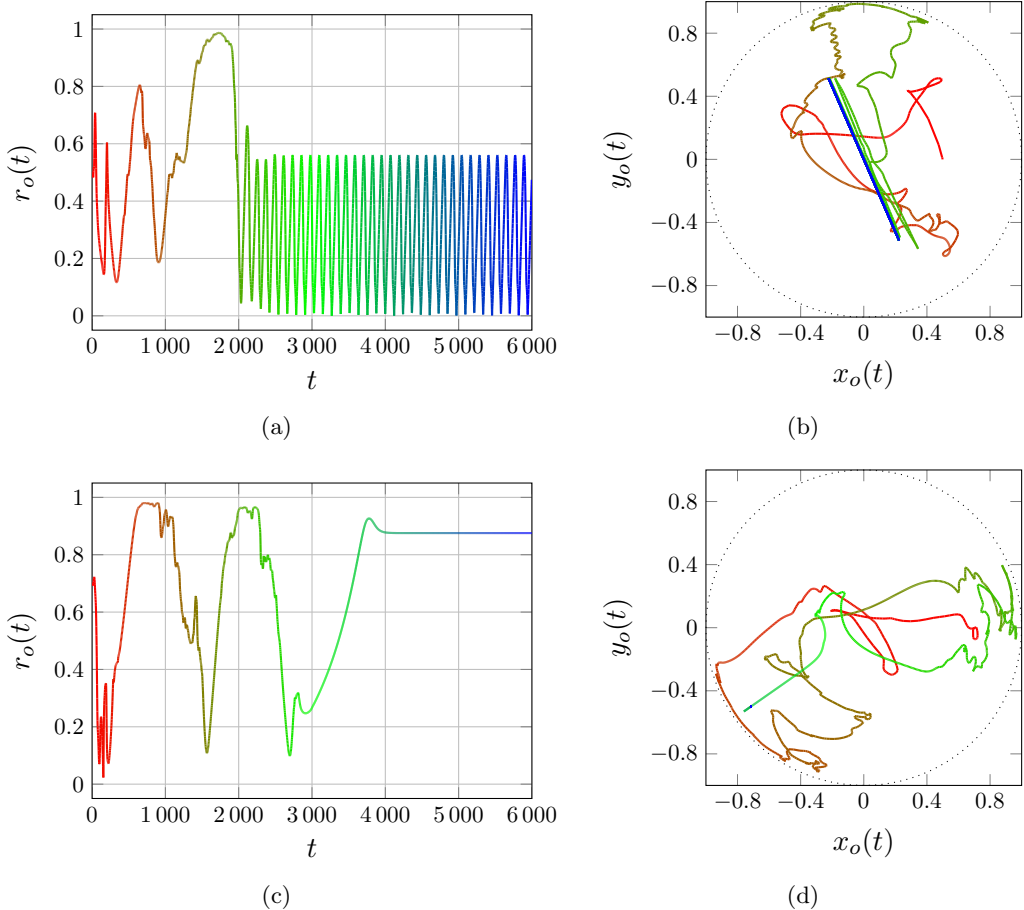
\begin{figure}
\begin{center}
  \subfigure[]{
    \begin{tikzpicture}
      \begin{axis}
        [ 
        ymin = -0.05,
        ymax = 1.05,
        xmin = 0,
        xmax = 6000,
        ylabel={$r_o(t)$},
        xlabel={$t$},
        tick scale binop=\times,
        width=0.45\textwidth,
        height=0.35\textwidth,
        grid=major,
        max space between ticks=25,
        ]
        \addplot+[no marks, thick, line join=round, , mesh, point meta=\thisrowno{0}, colormap/bluered,
        colormap={}{ %
           [1cm] color(0cm)=(red) color(1cm)=(green) color(2cm)=(blue)
        }] table[x
        expr=\thisrowno{0}, y expr=\thisrowno{8}, col sep=space] {Figures/pp-example-limit-e10.dat};
      \end{axis}
    \end{tikzpicture}
  }
  \subfigure[]{
    \begin{tikzpicture}
      \begin{axis}
        [ 
        ymin = -1,
        ymax = 1,
        xmin = -1,
        xmax = 1,
        xlabel={$x_o(t)$},
        ylabel={$y_o(t)$},
        tick scale binop=\times,
        width=0.35\textwidth,
        height=0.35\textwidth,
        xtick={-0.8,-0.4,0.0,0.4,0.8},
        ytick={-0.8,-0.4,0.0,0.4,0.8},
        ]
        \addplot+[no marks, thick, mesh, point meta=\thisrowno{0}, colormap/bluered,
        colormap={}{ %
           [1cm] color(0cm)=(red) color(1cm)=(green) color(2cm)=(blue)
        }] table[x
        expr=\thisrowno{2}, y expr=\thisrowno{3}, col sep=space] {Figures/pp-example-limit-e10.dat};
        \addplot [samples=100, domain=0:2*pi,dotted] ( {cos(deg(x))}, {sin(deg(x))} );
      \end{axis}
    \end{tikzpicture}
  }
  \medskip
 \subfigure[]{
    \begin{tikzpicture}
      \begin{axis}
        [ 
        ymin = -0.05,
        ymax = 1.05,
        xmin = 0,
        xmax = 6000,
        ylabel={$r_o(t)$},
        xlabel={$t$},
        tick scale binop=\times,
        width=0.45\textwidth,
        height=0.35\textwidth,
        grid=major,
        max space between ticks=25,
        ]
        \addplot+[no marks, thick, line join=round, mesh, point meta=\thisrowno{0}, colormap/bluered,
        colormap={}{ %
           [1cm] color(0cm)=(red) color(1cm)=(green) color(2cm)=(blue)
        }] table[x
        expr=\thisrowno{0}, y expr=\thisrowno{8}, col sep=space] {Figures/pp-example-fixed-e10.dat};
      \end{axis}
    \end{tikzpicture}
  }
  \subfigure[]{
    \begin{tikzpicture}
      \begin{axis}
        [ 
        ymin = -1,
        ymax = 1,
        xmin = -1,
        xmax = 1,
        xlabel={$x_o(t)$},
        ylabel={$y_o(t)$},
        tick scale binop=\times,
        width=0.35\textwidth,
        height=0.35\textwidth,
        xtick={-0.8,-0.4,0.0,0.4,0.8},
        ytick={-0.8,-0.4,0.0,0.4,0.8},
        ]
        \addplot+[no marks, thick, mesh, point meta=\thisrowno{0}, colormap/bluered,
        colormap={}{ %
           [1cm] color(0cm)=(red) color(1cm)=(green) color(2cm)=(blue)
        }] table[x
        expr=\thisrowno{2}, y expr=\thisrowno{3}, col sep=space] {Figures/pp-example-fixed-e10.dat};
        \addplot [samples=100, domain=0:2*pi,dotted] ( {cos(deg(x))}, {sin(deg(x))} );
      \end{axis}
    \end{tikzpicture}
  }
\caption{(a,c) Object center radial location $r_o(t)$ histories, and (b,d) $\bx_o(t) = \big(x_o(t),y_o(t)\big)$ trajectories for $\alpha = 4.8$ and $\zeta = 1.0$ for (a,b) $\bx_o(0) = (0.5,0)$, which leads to a limit cycle, and (c,d) $\bx_o = (0.7,0)$, which leads to a fixed point.  The same color pattern tracks time evolution in all frames.  Animations of these two simulations are shown in the supplementary material Movies~1 and 2.}\label{fig:basictraj}
\end{center}
\end{figure}

Figure~\ref{fig:basictraj} shows two example object trajectories for the circle-in-circle configuration.  They are for the same $\alpha = 4.8$, $\zeta = 1.0$, and $d_T = 0.01$ parameters, but with different initial positions of the object:  $x_o = 0.5$ versus $x_o=0.7$.  This difference seeds a significant divergence in subsequent trajectories, as would any perturbation of this chaotic system.  Both at first show a meandering behavior, with trajectories in figures~\ref{fig:basictraj} (b) and (d) often in nearly straight lines, consistent with some observations in experiments \cite{Ray:2023}.  The overall flow is characterized by the formation, interaction, and annihilation or absorption of multiple $\pm\shalf$ defects in the nematic order.  The local defect degree within a contour is defined in the usual way,
\begin{equation}
  m  = \frac{1}{\pi} \oint \theta \, ds,
  \label{e:m}
\end{equation}
for a contour around the potential defect location, $\theta$ being the nematic orientation angle, which is calculated based on the eigenvectors of $\bD$.  A radial arrangement, like a point electrical charge solution, would yield $m=\pm 1$ but is not observed in the fluid, presumably because it is so energetically unfavorable.  For a non-polar suspension, the nematic arrangement can also make a half twist around the closed contour, yielding $m=\pm\shalf$, which are commonly observed.

\begin{figure}
  \begin{center}
    \subfigure[$t=2752$ to $2772$]{ \includegraphics[width=0.32\textwidth]{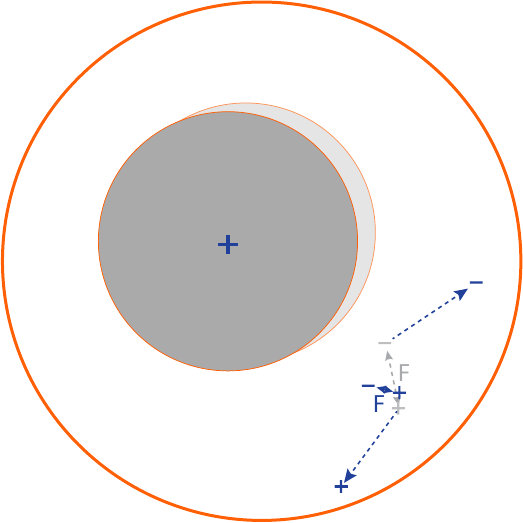}}
    \subfigure[$t=2772$ to $2781$]{    \includegraphics[width=0.32\textwidth]{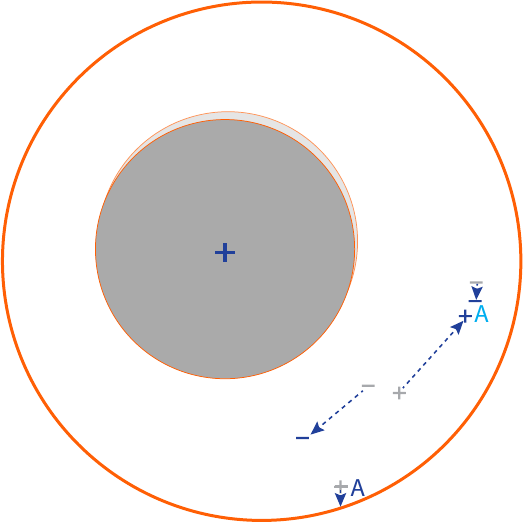}}
    \subfigure[$t=2782$ to $2816$]{    \includegraphics[width=0.32\textwidth]{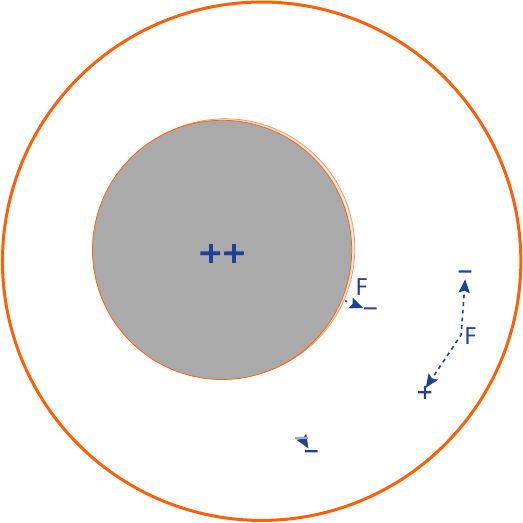}}
    \subfigure[$t=2816$ to $2818$]{    \includegraphics[width=0.32\textwidth]{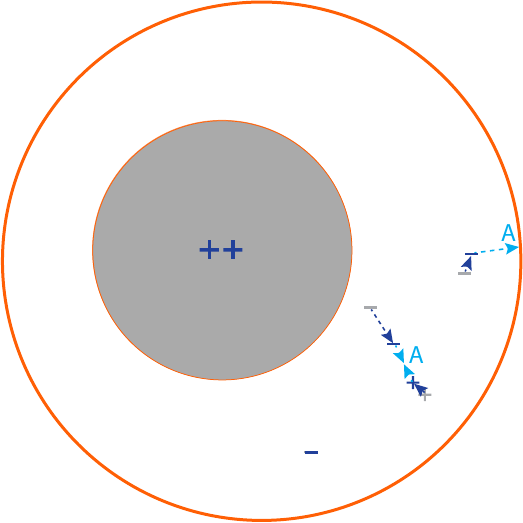}}
    \subfigure[$t=2826$ to $2999$]{    \includegraphics[width=0.32\textwidth]{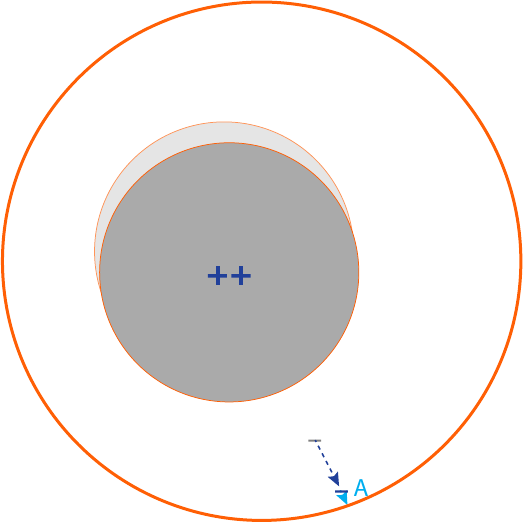}}
  \end{center}
  \caption{Trajectories of the defects leading to the final $m_\circ=+1$ state, and eventually the fixed point.  The $+$ and $-$ symbols indicate $m=\pm \shalf$ nematic defect.  Each frame represents locations at the two times labeled, with arrows indicating the displacement of the defects in this period.  Greys indicate earlier times, blue indicates the current time, and light blue indicates events about to happen.  The symbol A indicate annihilation of a pair or absorption to a wall boundary, and F indicates formation.  The lighter gray circle indicates the object at the earlier time. }\label{fig:sing-motion}
\end{figure}

The observed defects form individually on the object or container wall or as oppositely signed pairs.  The object itself also carries a nematic charge $m_\circ$, which is quantified by applying the same integral (\ref{e:m}) around its circumference.  Figure~\ref{fig:sing-motion} shows the evolution of the final defects and object $m_\circ$ for the case of figures~\ref{fig:basictraj} (c) and (d).  This particular series starts with $m_\circ= +\shalf$.  Subsequently, for the history considered in the figure, 3 pairs of $m\pm \shalf$ defects are seen to form in the bulk and one $m=-\shalf$ forms on the object.  Two pairs of opposite signed defects, each from different origins, annihilate and three are absorbed to the outer wall, leaving the object with $m_\circ = +1$.   (Not shown is a fourth pair that appears only briefly for 2 time units before self-annihilating.)  A similarly chaotic seeming pattern (not shown) leads to the $m_\circ=0$ condition for the case with limit-cycle behavior in figure~\ref{fig:basictraj} (a) and (b).  Defects are observed to form within or on the boundary of the less confined regions of the fluid, where the flow is more vigorous than in the more viscously constrained narrower region.  In the long time series observed, we only observed $\pm \shalf$ defects.   Only the object is observed to achieve a $m_\circ=\pm 1$ value.  

For large positive $\alpha$, the chaotic trajectories persist, seemingly indefinitely.  Some of these seem to bring the object in near contact ($\delta \lesssim 0.02$) with the wall, but it is always observed to be soon lifted away from the wall by the activity.   None of the cases presented required a contact-preventing constraint.  Decreasing $\alpha$ slowly causes flow to cease for $\alpha \to 0$, though it becomes unstable for $\alpha < 0$, as expected for extensors \cite{Ezhilan:2013}.  For the circle-in-circle geometry in this parameter range, the long-time outcome is either a fixed-point behavior, where the object becomes stationary near the boundary as in figure~\ref{fig:basictraj} (c) and (d) despite a sustained flow, or a limit-cycle behavior, with the object oscillating indefinitely on the scale of the container as in figures~\ref{fig:basictraj} (a)  and (b).  

\section{Fixed-point Behavior}
\label{s:fixed-point}

\subsection{Flow characteristics}

\begin{figure}
  \begin{center} \subfigure[Streamfunction]{\includegraphics[width=0.48\textwidth]{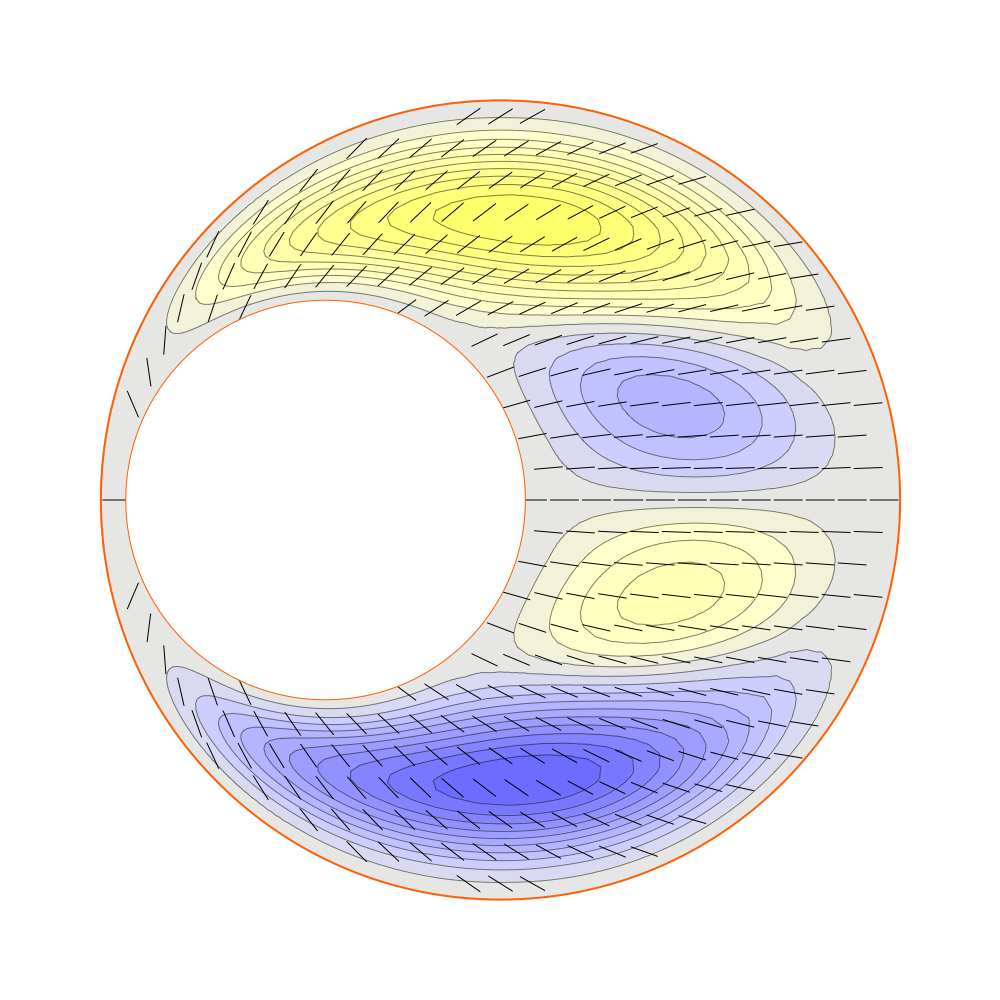}} \subfigure[Pressure]{\includegraphics[width=0.48\textwidth]{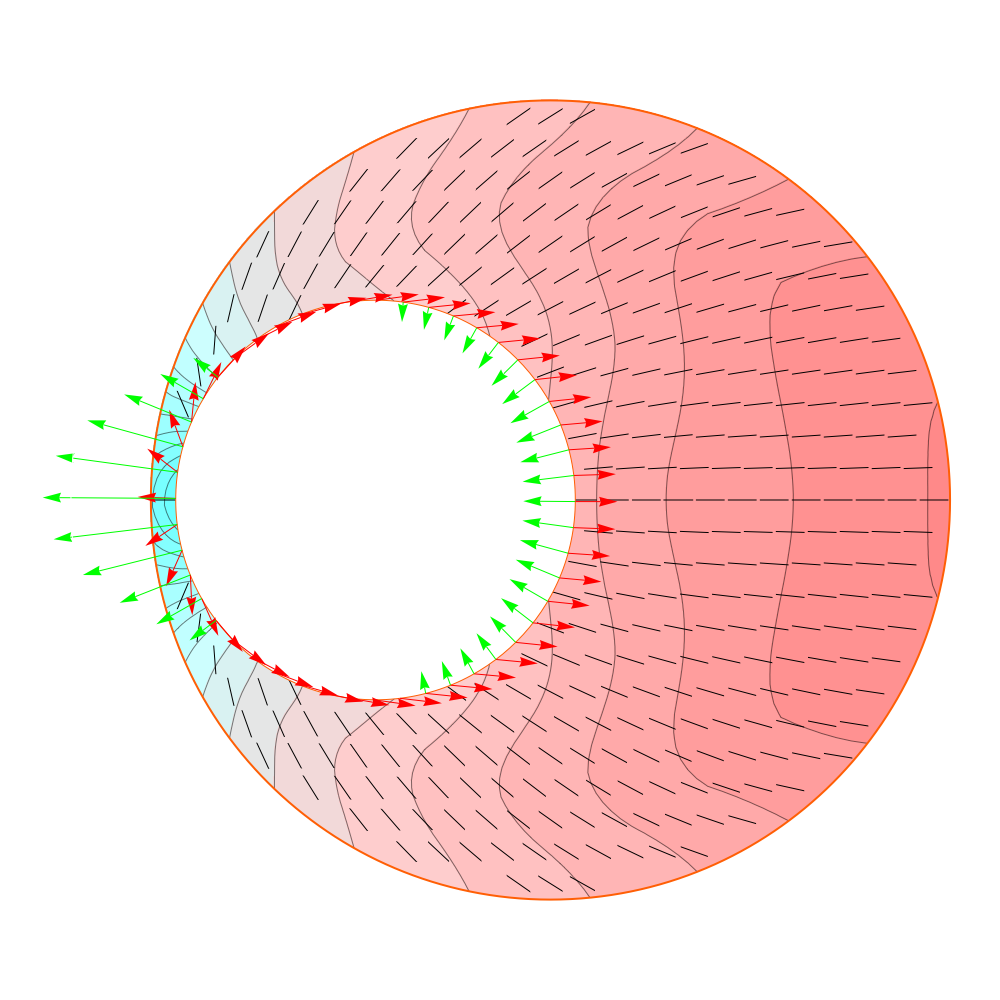}}
  \end{center}
  \caption{(a) Streamfunction and (b) pressure (and nematic orientation) visualization for $\alpha = 4.8$, $\zeta = 1.0$ for the fixed-point behavior for the case in figure~\ref{fig:basictraj} (c,d).  The contour levels are equally spaced between (a) $\psi \pm 0.0026$ and (b) $p = \pm 7.42$, which are based on the domain maximum $|\psi|$ and $|p|$, respectively.  The red colors in (b) are positive for pressure, and the undetermined constant is set such that the mean pressure on the object surface is zero.  Yellow contours in (a) indicate counter-clockwise flow.  The domain is rotated such that the object is at $y_o=0$, as shown.  The actual location is seen in figure~\ref{fig:basictraj} (d).  The arrows in (b) visualize the surface tractions:  red is the deviatoric $\alpha (\bD-\bI/2)\cdot\bn$ active traction and green is the $(-p\bI + 2 \bE)\cdot\bn$ hydrodynamic traction.}\label{fig:viz-fixed}
\end{figure}
\begin{figure}
  \begin{center}
    \subfigure[$t = 200$ ($r_o = 0.303$)]{\includegraphics[width=0.32\textwidth]{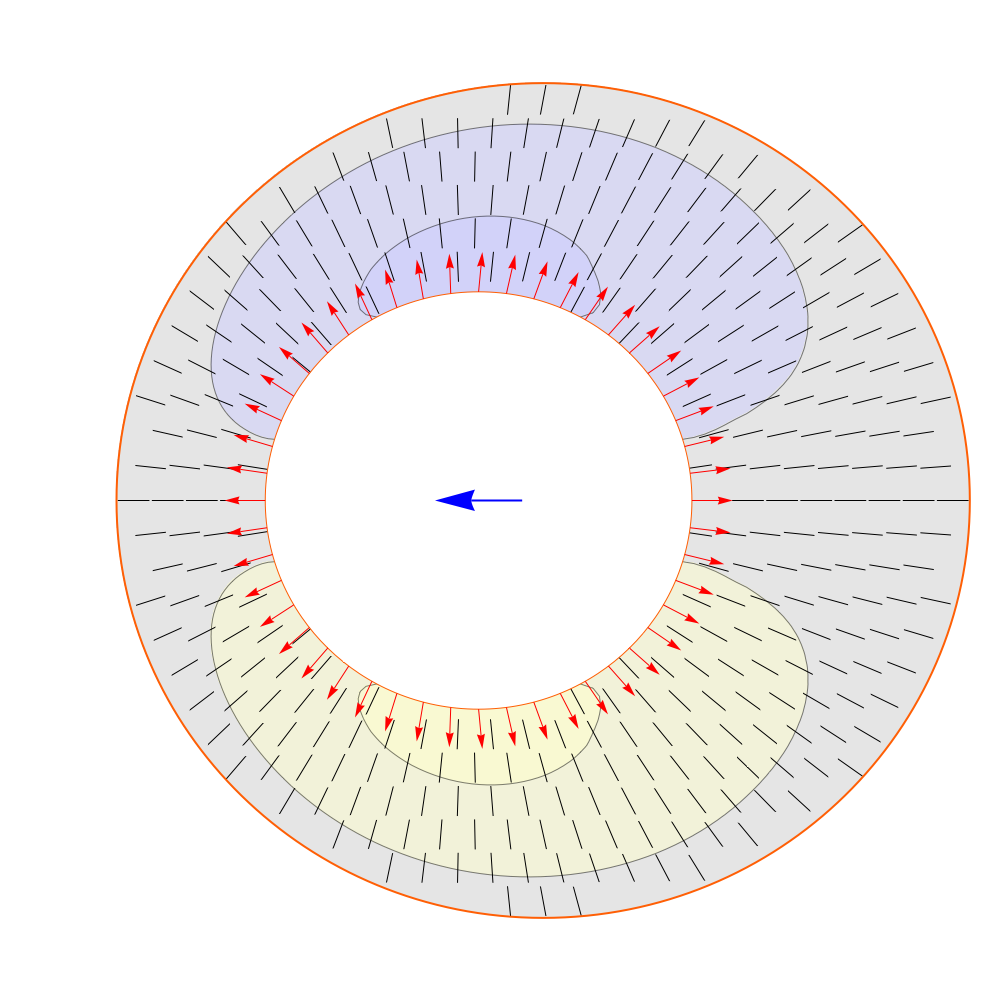}}
    \subfigure[$t = 760$ (maximum $-U_r$)]{\includegraphics[width=0.32\textwidth]{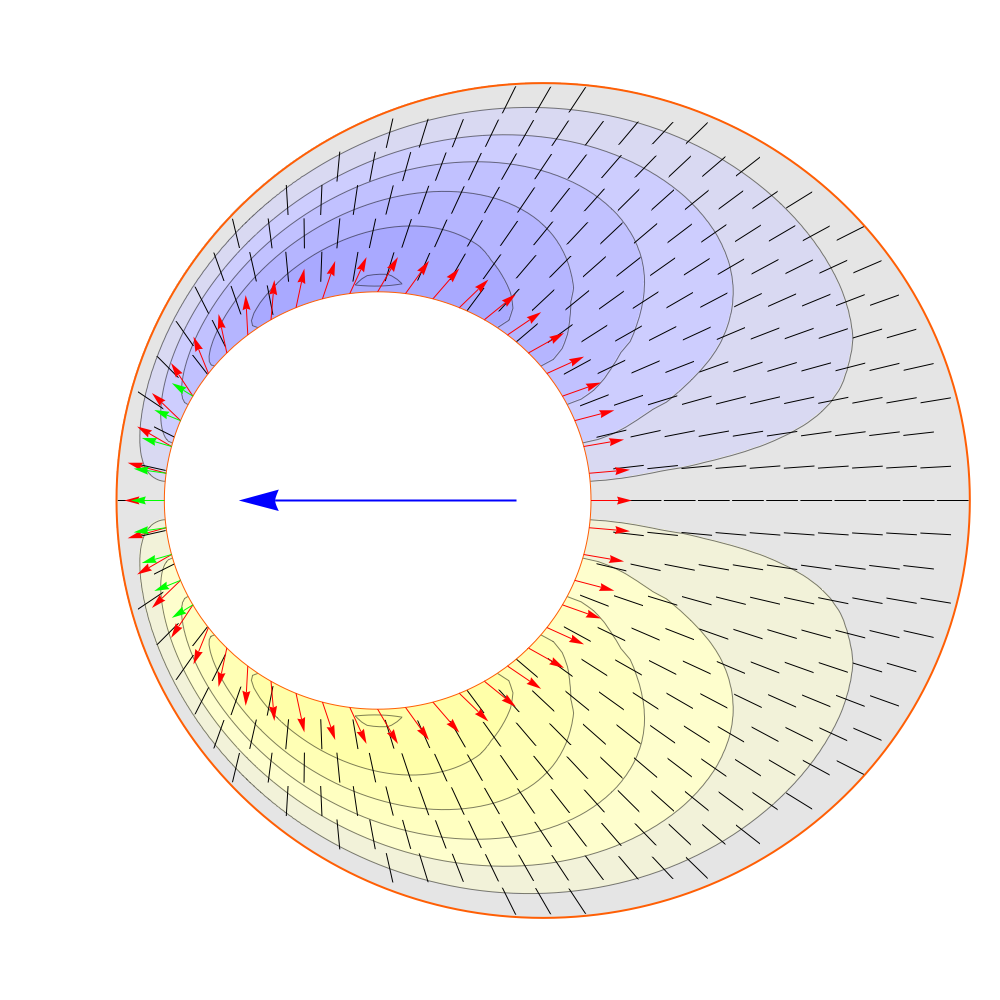}}
    \subfigure[$t = 820$ (first $r_o = r_o^\text{equilib.}$)]{\includegraphics[width=0.32\textwidth]{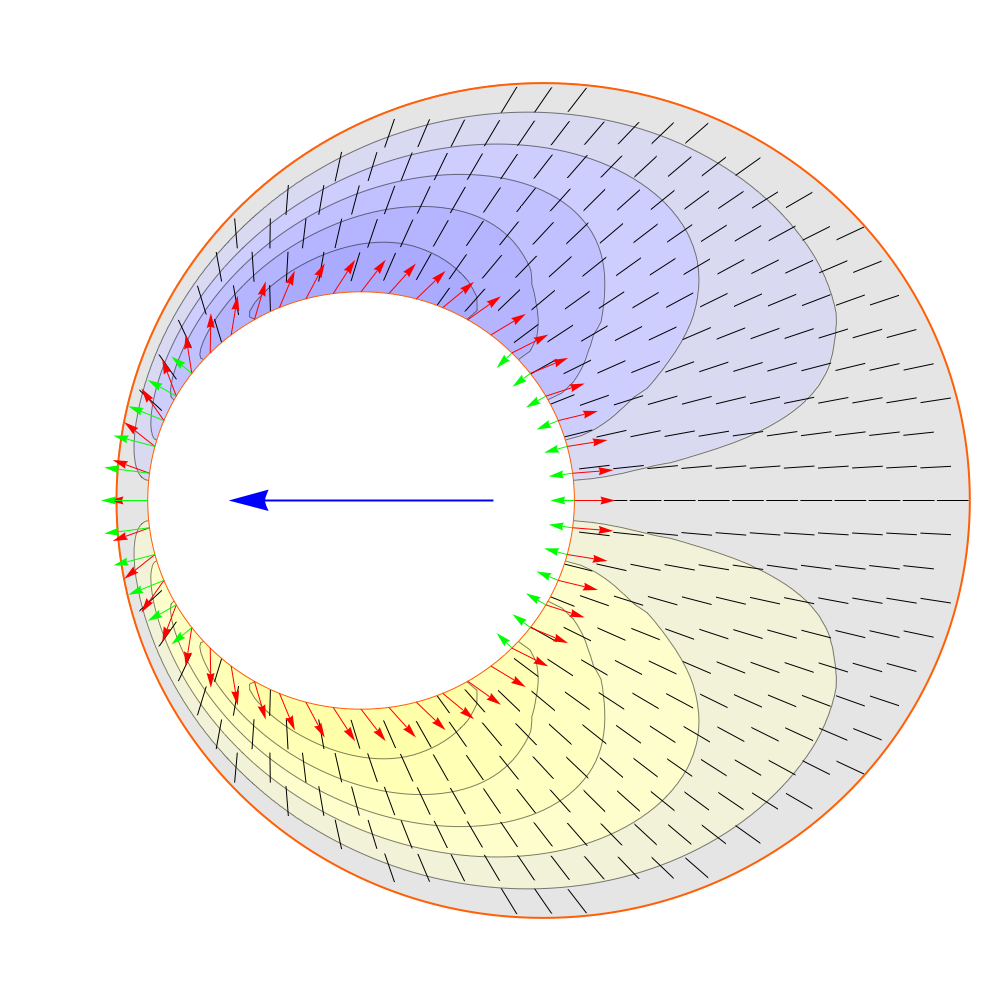}}\\
    \subfigure[$t = 880$ (just prior to reversal)]{\includegraphics[width=0.32\textwidth]{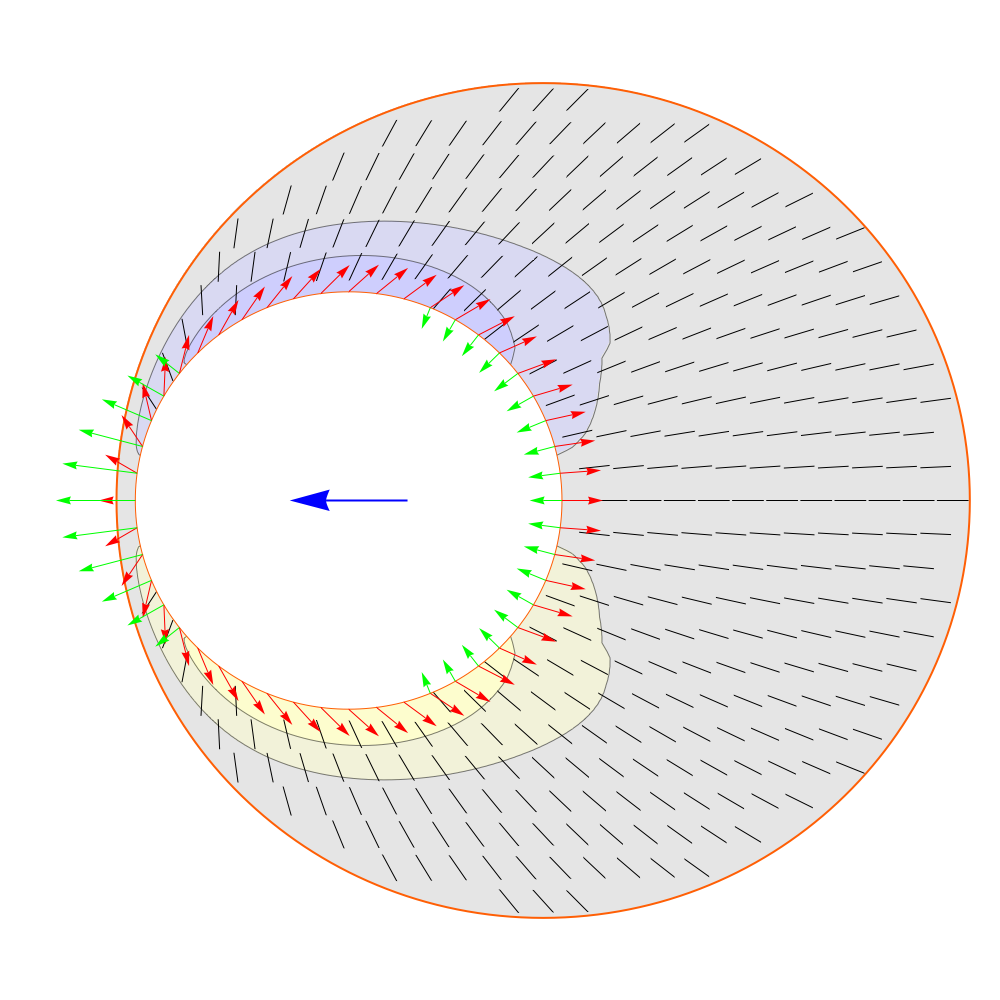}}
    \subfigure[$t = 960$ (maximum $+U_r$)]{\includegraphics[width=0.32\textwidth]{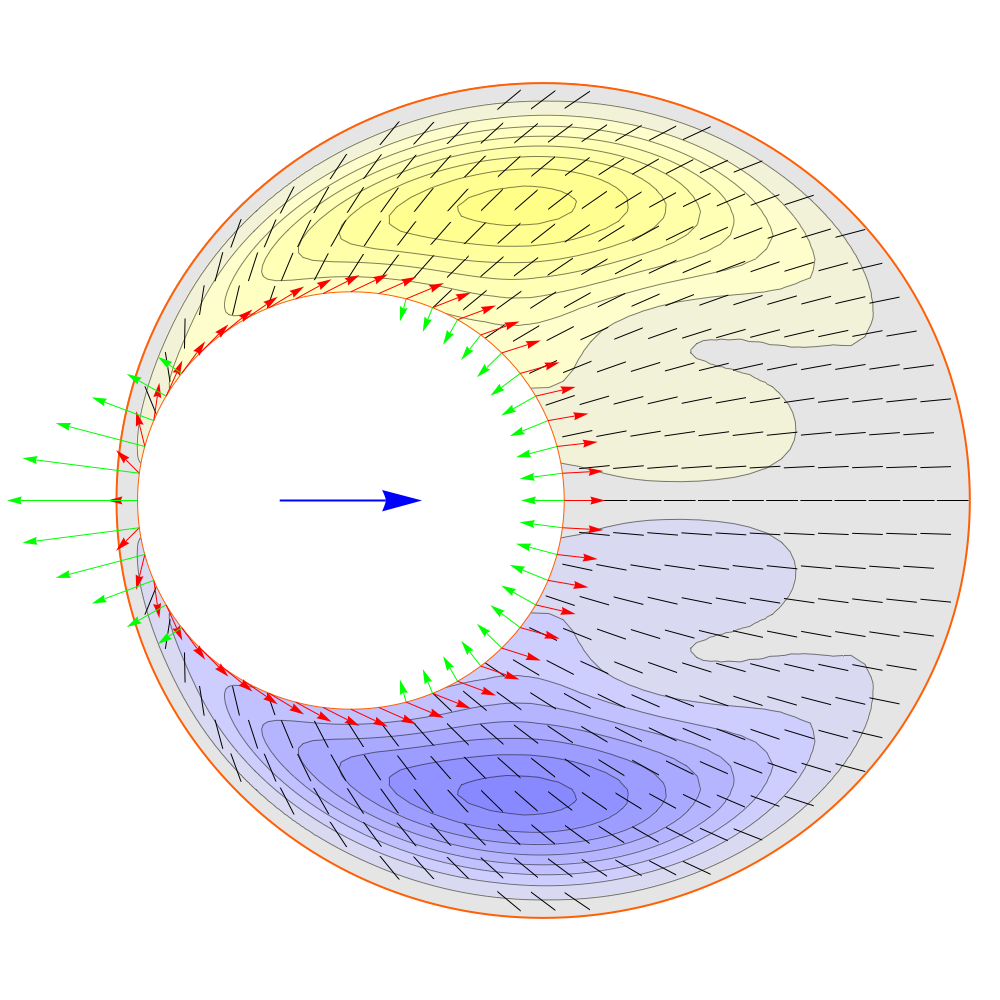}}
    \subfigure[$t \gtrsim 1200$ ($r_o = r_o^\text{equilib.}$, $U_r = 0$)]{\includegraphics[width=0.32\textwidth]{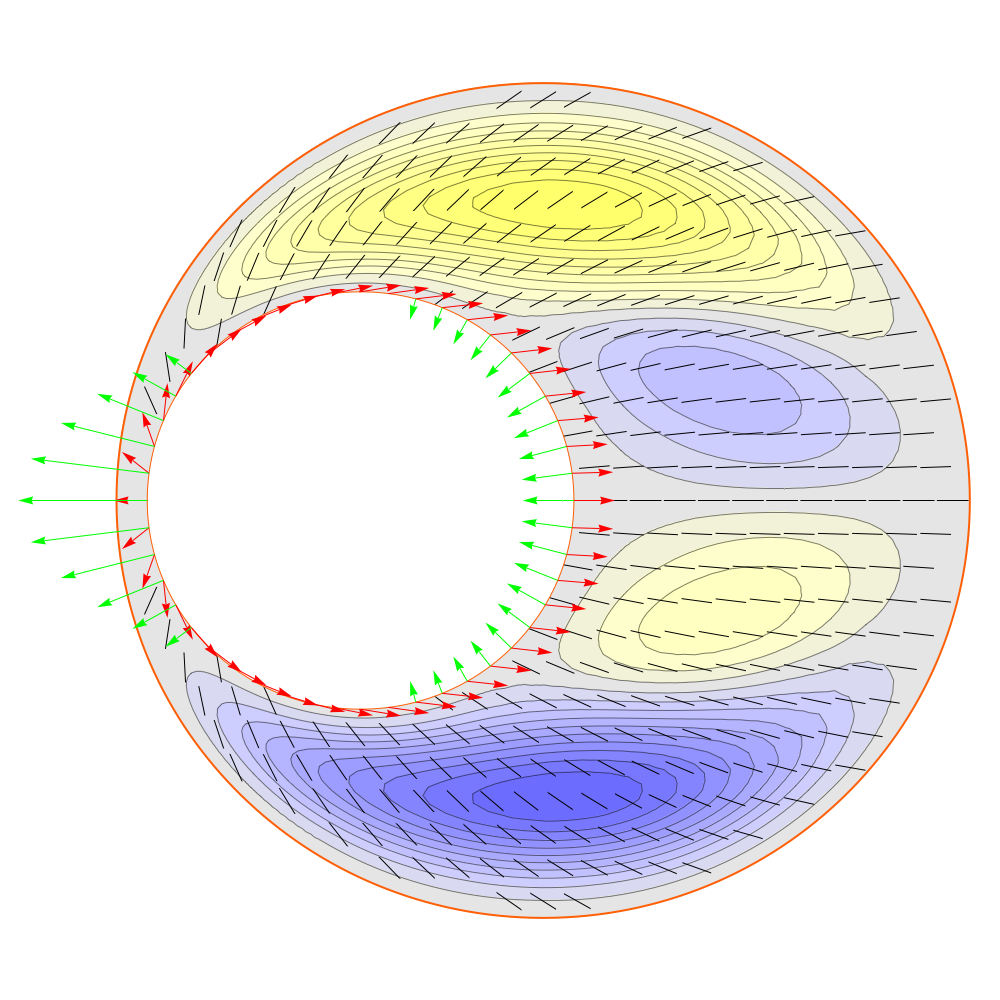}}
  \end{center}
  \caption{Streamfunction visualizations for $\alpha = 5$, $\zeta = 0.7$ for approach to a fixed point for the case of figure~\ref{fig:basictraj} (c,d).  The 20 streamfunction levels are equally spaced between $\psi \pm 0.0025$. The arrows visualize the surface tractions:  red is the active deviatoric factor $\alpha(\bD-\bI/2)\cdot\bn$ and green is the net hydrodynamics traction $(-p\bI + 2 \bE)\cdot\bn$.  The blue object velocity arrow length is proportional to its speed.}\label{fig:viz-approach-fixed}
\end{figure}

Figure~\ref{fig:viz-fixed} visualizes the case of figure~\ref{fig:basictraj} (c,d).  In all cases observed, the object approaches the fixed point via a wall-normal trajectory.  However, its approach is not a sequence of quasi-steady conditions, as might be expected for Newtonian-fluid Stokes flow.  It instead depends on the hydrodynamics coupled with the organized activity of the fluid.  Figure~\ref{fig:fixedpoint-approach} shows the velocity of the object as it approaches the fixed point, and figure~\ref{fig:viz-approach-fixed} visualizes the corresponding flow.  In this case, the domain was initialized with an axisymmetric (about the container center) radial nematic perturbation with $\bD$ eigenvalues $\lambda_1 = 0.499$ and $\lambda_2 =0.501$, which develops rapidly into a field with nearly constant $\lambda_1 \in [0.016,0.026]$ (and $\lambda_2 = 1-\lambda_1$) by time $t=2$, consistent with the fast alignment time scale $\tau_\zeta$ in table~\ref{tab:taus}.  The object moves only from $r_o = 0.3$ to $ r_o= 0.292$ in this transient period, consistent with the slow object advective time scale $\tau_{u_\circ}^\ell$ (table~\ref{tab:taus}), and it retains its initial  $m_\circ = +1$ nematic structure.  Experimentation identified that this constructed case with the initial object center at $\bx_o = (-0.3,0)$ leads to immediate approach to the fixed point position at $\bx_o = (-0.855,0)$, matching the final post-chaos trajectory in figure~\ref{fig:basictraj} (c) and (d), but avoiding the chaotic wandering that proceeds it.

A radial overshoot of the fixed point, to $r_o = 0.92$, is visible.  This is somewhat surprising because for a fixed object, with $\bU = 0$ by constraint, the suspension for $r_o \ge 0.086$ is found to be unstable, which can be inferred from the fluctuating wall-normal net forces for fixed $r_o \in \{ 0.080, 0.0850, 0.0853, 0.0857, 0.0860\}$ cases also shown in figure~\ref{fig:fixedpoint-approach}.  The large force fluctuations on the fixed object at $r_o = 0.086$ persist in a seemingly statistically stationary way for very long $t=10^5$ simulations.  The wall-normal force samples shown cover only part of this time series; the wall-tangent component has similar fluctuations, with both positive and negative values.  The other four fixed-object cases show how the forces on the fixed objects are consistent with the slow, $U_r$-specified cases also shown.  For these smaller $r_o$ values, the radial nematic ordering leads to steady symmetric circulations and a net force toward the wall. 

The instability that arises for fixed $r_o\gtrsim 0.86$ in figure~\ref{fig:viz-approach-fixed} is slow to develop, and it does not appear until near contact for $U_r = 0.001$, which is close to observed wall-ward object speed for the free object in its approach.  This is demonstrated in figure~\ref{fig:fixedpoint-approach} for cases in which the object is moved at steady $U_r$ speeds.  All approach speeds have a region, closer to the wall than the static-object instability threshold around $r_o \approx 0.86$, for which motion is countered by a strong repulsion.  This region is broader for faster motion, and its time scale is consistent with the slower flow time scale $\tau_{u_\circ}^\ell$ in table~\ref{tab:taus}, which can be anticipated to be central to the hydrodynamic instabilities.  This meta-stable response returns the object from the long-time unstable near-wall region to its stable fixed point before instabilities grow to be consequential.

\begin{figure}
  \begin{center}
    \begin{tikzpicture}
      \begin{axis}
        [    
        axis y line*=right,
        axis x line = none,
        ymin = -15,
        ymax = 20,
        xmin = 0.798,
        xmax = 0.95,
        ylabel=$F_r$,
        yticklabel pos=right,
        width=0.65\textwidth,
        height=0.45\textwidth,
        unbounded coords=jump,
        ]
        \addplot [samples=100, domain=0:1, color=gray] ( x, 0 );
        \addplot [samples=100, domain=-20:5, color=gray, dashed] ( 0.85527, x);
        \node at (axis cs:0.84, 1.5) {{\footnotesize \color{orange} $U_r = 0 \; \to$}};
        \node at (axis cs:0.87, 18) {{\footnotesize \color{blue!25!white} $U_r = 3 \times 10^{-5}$}};
        \node at (axis cs:0.91, 17) {{\footnotesize \color{blue!50!white} $1 \times 10^{-4}$}};
        \node at (axis cs:0.934, 13.5) {{\footnotesize \color{blue!75!white} $3 \times 10^{-4}$}};
        \node at (axis cs:0.935, 1.3) {{\footnotesize \color{blue} $1 \times 10^{-3}$}};
        \node at (axis cs:0.85, 7.5) {{\begin{minipage}{0.3\textwidth}\footnotesize \color{blue!66!white} Force $F_r$ for \\ $U_r = \text{constant}$ \\ (right axis $\rightarrow$)\end{minipage}}};
        \node at (axis cs:0.845, -8.95) {{\begin{minipage}{0.3\textwidth}\footnotesize \color{red} Speed $U_r$ for free floating ($F_r=0$) \\($\leftarrow$ left axis) \end{minipage}}};
        \addplot+[no marks, thin, color=blue!25!white, x filter/.expression={(x < 0.88 ? \pgfmathresult : NaN)}] table[x expr=\thisrowno{0}, y expr=\thisrowno{1}, col sep=space] {Figures/r-drag-prescribedU=0.00003-e50.dat};
        \addplot+[mark=*, blue!25!white, mark options={fill= blue!25!white,draw=none}, draw=none, mark repeat=1, only marks, mark size = 0.5pt, x filter/.expression={(x < 0.88 ? \pgfmathresult : NaN)}] table[x expr=\thisrowno{0}, y expr=\thisrowno{1}, col sep=space] {Figures/r-drag-prescribedU=0.00003-e100.dat};
        \addplot+[no marks, thin, color=blue!50!white, x filter/.expression={(x < 0.91 ? \pgfmathresult : NaN)}] table[x expr=\thisrowno{0}, y expr=\thisrowno{1}, col sep=space] {Figures/r-drag-prescribedU=0.0001-e10.dat};
        \addplot+[mark=*, blue!50!white, mark options={fill=blue!50!white,draw=none}, draw=none, mark repeat=1, only marks, mark size = 0.5pt, x filter/.expression={(x < 0.91 ? \pgfmathresult : NaN)}] table[x expr=\thisrowno{0}, y expr=\thisrowno{1}, col sep=space] {Figures/r-drag-prescribedU=0.0001-e100.dat};
        \addplot+[no marks, thin, color=blue!75!white, solid] table[x expr=\thisrowno{0}, y expr=\thisrowno{1}, col sep=space] {Figures/r-drag-prescribedU=0.0003-e10.dat};
        \addplot+[mark=*, blue!75!white, mark options={fill=blue!75!white,draw=none}, draw=none, mark repeat=1, only marks, mark size = 0.5pt] table[x expr=\thisrowno{0}, y expr=\thisrowno{1}, col sep=space] {Figures/r-drag-prescribedU=0.0003-e100.dat};
        \addplot+[no marks, thin, color=blue, solid] table[x expr=\thisrowno{0}, y expr=\thisrowno{1}, col sep=space] {Figures/r-drag-prescribedU=0.001-e10.dat};
        \addplot+[mark=*, blue, mark options={fill=blue,draw=none}, draw=none, mark repeat=10, only marks, mark size = 0.5pt] table[x expr=\thisrowno{0}, y expr=\thisrowno{1}, col sep=space] {Figures/r-drag-prescribedU=0.001-e10.dat};
        \addplot+[mark=x, orange, mark options={fill=orange,draw=none}, draw=none, mark repeat=1, only marks, mark size = 2.5pt] table[x expr=\thisrowno{0}, y expr=\thisrowno{1}, col sep=space] {Figures/d-v-r-U=0.dat};
        \addplot+[mark=x, orange, mark options={fill=orange,draw=none}, draw=none, mark repeat=1, only marks, mark size = 0.8pt] table[x expr=\thisrowno{0}, y expr=\thisrowno{1}, col sep=space] {Figures/d-samp.dat};
      \end{axis}
      \begin{axis}
        [ 
        ymin = -0.0015,
        ymax = 0.002,
        xmin = 0.798,
        xmax = 0.95,
        ylabel={$U_r$},
        xlabel={$r_o$},
        tick scale binop=\times,
        width=0.65\textwidth,
        height=0.45\textwidth,
        xtick=\empty,
        xtick={0.8, 0.85, 0.85527, 0.9, 0.95},
        xticklabels={0.8,, 0.855, 0.9, 0.95},
        ]
        \addplot+[no marks, thick, color=red] table[x
        expr=\thisrowno{0}, y expr=\thisrowno{1}] {Figures/r-Ux-a5-z0.7.dat};
        \addplot+[mark=*, red, mark options={fill=white,draw=none}, draw=none, mark repeat=10, only marks, mark size = 0.7pt] table[x expr=\thisrowno{0}, y expr=\thisrowno{1}, col sep=space] {Figures/r-Ux-a5-z0.7.dat};
      \end{axis}
    \end{tikzpicture}
  \end{center}
  \caption{For $\alpha = 5$, $\zeta = 0.7$, curves show the free floating normal speed from the container wall $U_r$ at it approaches the $r_o = 0.855$ fixed point and the net wall-normal active stress forces $F_r = -\bn_w \cdot \nabla \cdot \bsigma_a$ for prescribed $U_r$ as labeled, including five fixed ($U_r=0$) cases indicated with $\times$ symbols, only one of which (for $r_o = 0.86$) is unsteady after initial transients.  As defined, $F_r$ are taken as positive when directed away from the container wall.  All the plotted symbols are equally spaced in time with $\Delta t = 10$.   The motion in all cases is normal to the wall, either fixed that way for specified $U_r$ or a result of the dynamics for the free object.  }\label{fig:fixedpoint-approach}
\end{figure}

A $m_\circ=+1$ object can also have an associated circumferential nematic structure.  Initializing this case also yields a radial trajectory toward the container wall, but it does not then end up in a stable position.  Rather than reversing gently and arriving at the stable point, it repels rapidly to near the center of the container, inducing disclinations that interact chaotically.  Ultimately, as for any case observed, depending on the chaotic dynamics, this case falls into either the similar fixed-point or limit-cycle as discussed in section~\ref{s:limit-cycle} Since $m=+\shalf$ disclinations are polar, initializing the object with $m_\circ=+\shalf$ initiates quick motion toward a wall, but this generates more defects and a period of chaotic motion before again arriving at either the fixed-point or limit-cycle solutions.  Initializing with $m_\circ=-\shalf$ has this same outcome, though chaos arises slowly since the circle is initially in a balanced configuration.

\subsection{Parametric dependence}
\label{s:prms-fixed-point}

\begin{figure}
\begin{center}
  \subfigure[]{
    \begin{tikzpicture}
      \begin{axis}
        [ 
        ymin = 0.815,
        ymax = 0.865,
        xmin = 0,
        xmax = 6,
        ylabel={$r_o$},
        xlabel={$\alpha$},
        tick scale binop=\times,
        width=0.45\textwidth,
        height=0.35\textwidth,
        grid=major,
        max space between ticks=25,
        ]
        \addplot+[no marks, thick, color=red] table[x
        expr=\thisrowno{0}, y expr=\thisrowno{1}, col sep=space] {Figures/alpha-r-e10000-fixed.dat};
      \end{axis}
    \end{tikzpicture}
  }
  \subfigure[]{
    \begin{tikzpicture}
      \begin{axis}
        [ 
        ymin = 0.8,
        ymax = 0.9,
        xmin = 0,
        xmax = 8.2,
        ylabel={$r_o$},
        xlabel={$\zeta$},
        tick scale binop=\times,
        width=0.45\textwidth,
        height=0.35\textwidth,
        grid=major,
        max space between ticks=25,
        ]
        \addplot+[no marks, thick,color=green!80!black] table[x
        expr=\thisrowno{0}, y expr=\thisrowno{1}, col sep=space] {Figures/zeta-r-fixed.dat};
      \end{axis}
    \end{tikzpicture}
  }
  \subfigure[]{
    \begin{tikzpicture}
      \begin{axis}
        [ 
        ymin = 0.81,
        ymax = 0.88,
        xmin = 0.008,
        xmax = 0.0117,
        ylabel={$r_o$},
        xlabel={$d_T$},
        tick scale binop=\times,
        width=0.45\textwidth,
        height=0.35\textwidth,
        grid=major,
        max space between ticks=25,
        ]
        \addplot+[no marks, thick,color=blue] table[x
        expr=\thisrowno{0}, y expr=\thisrowno{1}, col sep=space] {Figures/dT-r-fixed.dat};
      \end{axis}
    \end{tikzpicture}
  }
\caption{Equilibrium radius dependence on (a) $\alpha$, (b) $\zeta$, (c) $d_T$ for a case initiated with $\alpha = 5$, $\zeta = 0.7$, and $d_T = 0.01$.}\label{fig:fixed-prms}
\end{center}
\end{figure}
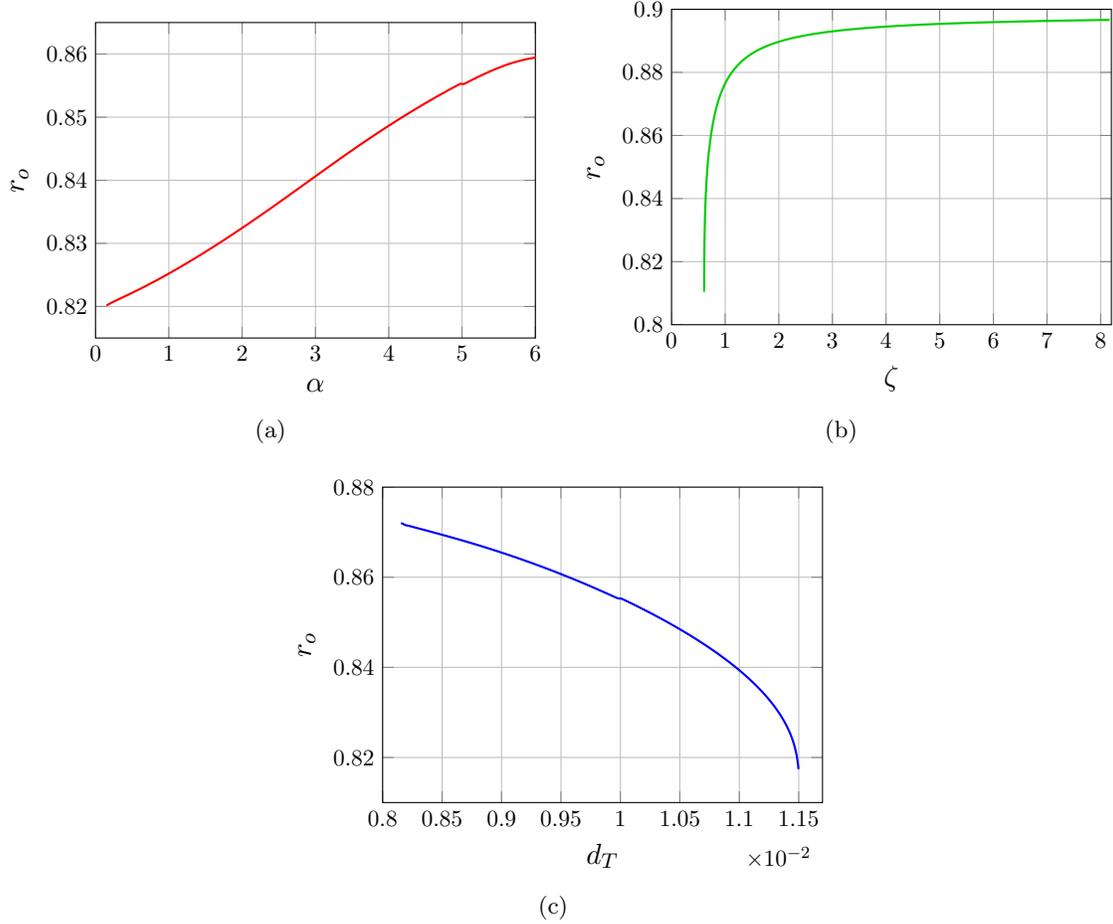

The basic form of the fixed-point solution is relatively insensitive to changes in the parameter values as they are slowly adjusted from a $\alpha = 5$, $\zeta = 0.7$, $d_T=0.01$ starting point. Figure~\ref{fig:fixed-prms} (a) shows the effect of changing $\alpha$.  The existence of the fixed point configuration extends to $\alpha$ near zero, where the circulations are similar to those visualized in figure~\ref{fig:viz-fixed} (a), though much weaker.  Flow eventually ceases as $\alpha \to 0$.  For negative $\alpha$, the suspension is unstable, as anticipated in the stability analysis of a similar kinetic model \cite{Ezhilan:2013}.  Near $\alpha = 6$, instabilities seem to arise spontaneously in the larger space, consistent with the expectation of the suspension itself becoming unstable for wave lengths that decrease with increasing $\alpha$. The resulting flow structures are observed to move the object away from its position near the wall.  

Varying $\zeta$ has a different outcome, as seen in figure~\ref{fig:fixed-prms} (b).  Increasing $\zeta$ leads $r_o^{\text{eq}}$ to approach an apparent limit near 0.9.  This is consistent with reaching a fully aligned limit of the nematic.  At this highest $\zeta =8$, the maximum of $\lambda_1$ is only $\lambda_1 = 0.007$, hence $\bD\cdot\bD - \bD\cddot\bS \approx 0$, and dependence on $\zeta$ in both (\ref{e:mom}) and (\ref{e:D}) is lost.  An important strength of the Bingham closure is its representation of this limit \cite{Gao:2017,Weady:2022}.  Decreasing $\zeta$ leads to failure of the fixed point into a chaotic flow for $\zeta \lesssim 0.6$.  In this case, it seems that the translational diffusion overwhelms the $\zeta$-driven formation of nematic gradients in the narrow region, consistent with the $\tau_\zeta / \tau_T^\delta \approx 1$ in table~\ref{tab:taus}.

The diffusion affords the narrowest range of fixed-point stability, as seen in figure~\ref{fig:fixed-prms} (c).  Lowering $d_T$ allows steeper $\bD$ gradients to exist, which in turn strengthens the flow and hydrodynamic attraction toward the wall.  However, below $d_T\approx 0.008$, the overall suspension becomes unstable.  As for increasing $\alpha$, it first shows fluctuations in the larger fluid region opposite the object.  In contrast, increasing $d_T$ decreases $r_o^{\text{eq}}$ (increasing the gap between the object and the wall), but this also causes the fixed-point stability to fail for large enough $d_T$.  The visualizations of the nematic alignment in figure~\ref{fig:viz-fixed} show that it varies significantly in the narrow gap, which creates a net tension on the fluid and a concomitant low pressure that pulls the object toward the container wall, countering the $\alpha \bD$ component that primarily pulls it in the opposite direction.  This is analyzed in the sharply aligned limit in the following section.  Similar to decreasing $\zeta$, increasing $d_T$ suppresses the gradients that are critical for maintaining this, which decreases the low-pressure attraction, and eventually the low pressure is insufficient to counter $\alpha \bD$.

\subsection{Sharply aligned nematic model}

In the fixed-point configuration for $\alpha = 5$, $\zeta = 0.7$, and $d_T=0.01$, the maximum value of the smaller $\bD$ eigenvalue is $\lambda_1 = 0.093$, and it is near its equilibrium value $\lambda_1 = 0.016$ in much of the domain.  Such strong alignment suggests perfect alignment as a candidate model for examining the fixed-point equilibrium structure.  This scenario can be represented as $\bD = \bP \bD_0 \bP^T$, where $\bP(\theta)$ is a rotation tensor
\begin{equation}
  \bP = \begin{pmatrix}
    \cos \theta & \sin \theta \\
    -\sin \theta & \cos \theta \end{pmatrix}
  \label{e:Rrot}
\end{equation}
and $\bD_o$ is the reference alignment
\begin{equation}
  \bD_o =
\begin{pmatrix}
    1 & 0 \\
    0 & 0
  \end{pmatrix}.
  \label{e:Do}
\end{equation}
The alignment energy leading to this fully aligned limit, specifically the phenomenological Maier--Saupe-like potential \cite{Ezhilan:2013} that leads to (\ref{e:mom}) and (\ref{e:D}) \cite{Gao:2017}, can be interpreted as having free energy $U \propto |\nabla \theta|^2$, for which a variational approach leads to a Laplace equation governing the orientation \cite{Chandler:2023}:
\begin{equation}
  \Delta \theta = 0.
  \label{e:Laptheta}
\end{equation}
The lack of wall-induced alignment in our description indicates
\begin{equation}
  \bn\cdot\nabla\theta = 0
  \label{e:thetabc}
\end{equation}
on the container at $|\bx| = r = R$ and on the object $|\bx-\bx_o| = a$.  Clearly, solutions of (\ref{e:Laptheta}) with (\ref{e:thetabc}) describe local energy minimizing configurations; constant-$\theta$ produces the minimum global energy.  

For a circle of radius $R_1$ centered within a circle of radius $R_2$ container, $m_\circ = +1$ radial nematic ordering has
\begin{equation}
  \theta(r,\phi) = \phi   \qquad \text{for}\quad R_1< r < R_2 \quad\text{and}\quad 0 \le \phi < 2 \pi,
  \label{e:wsol}
\end{equation}
where $(r,\phi)$ are container-centered cylindrical coordinates, with corresponding Cartesian coordinates $\xi=r\cos\phi$ and $\eta = r\sin\phi$.
A conformal mapping from this concentric circle configuration in the $w = \xi + i \eta$ complex plane to non-concentric circles in the $z = x + i y$ plane is \cite{Kober:1957}
\begin{equation}
  w(z) = t\, \frac{R_1}{r_1} \frac{(z-d) + s }{(z-d)+t }.
  \label{e:mapping}
\end{equation}
Following from (\ref{e:wsol}), this gives the solution $\theta(z) = \arg w(z)$.  The mapping parameters $s$ and $t$ in (\ref{e:mapping}) solve \cite{Kober:1957}
\begin{equation}
s t = r_1^2 \qquad\text{and}\qquad (d - s) (d - t) = r_2^2,
\end{equation}
and
the relationship between the mapped concentric circle radii $R_1$ and $R_2$ and the non-concentric $d$-offset circles of radii $r_1$ and $r_2$ is
\begin{equation}
  R_2 = R_1 \frac{r_2}{r_1} \left| \frac{t}{d - t}\right|.
\end{equation}
The smaller $r_1=a$ circle (the object) is centered at $z=0$ and the larger $r_2=R$ circle is centered at $(d,0)$.  

\begin{figure}
  \begin{center} \subfigure[Streamfunction]{\includegraphics[width=0.48\textwidth]{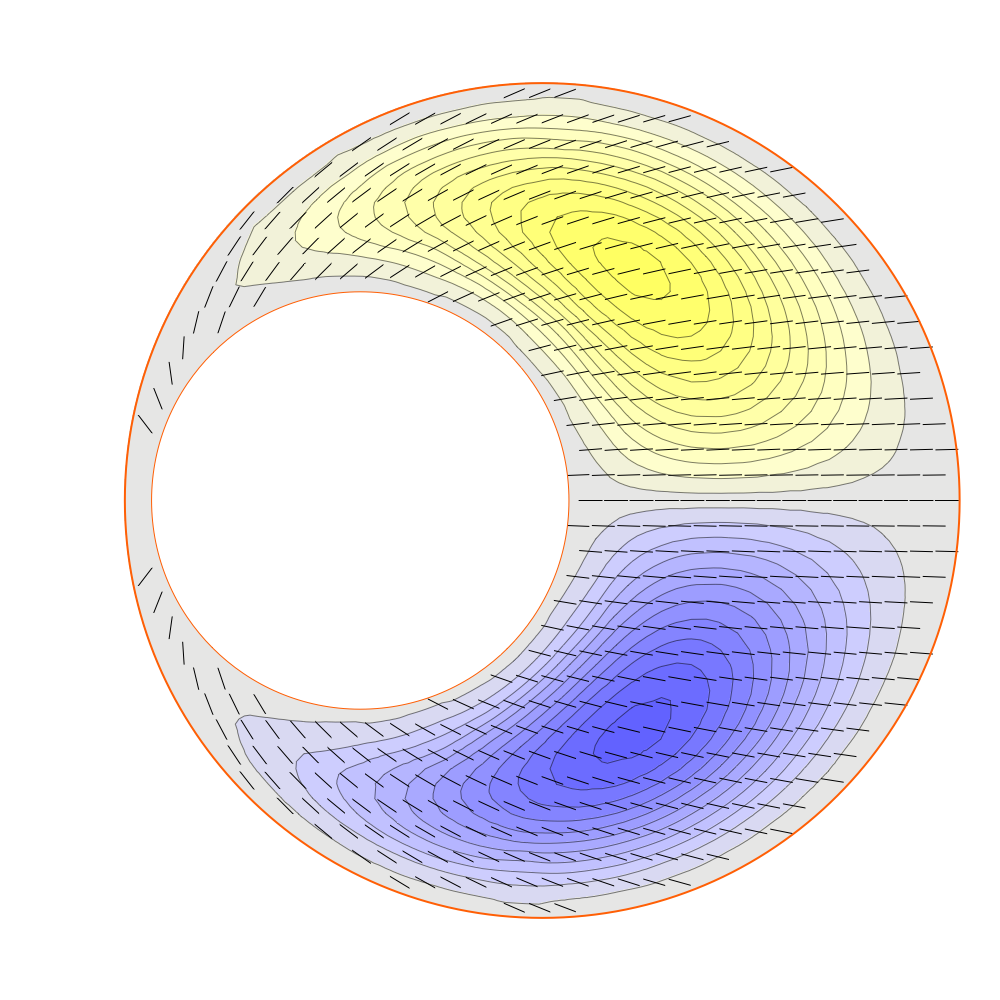}} \subfigure[Pressure]{\includegraphics[width=0.48\textwidth]{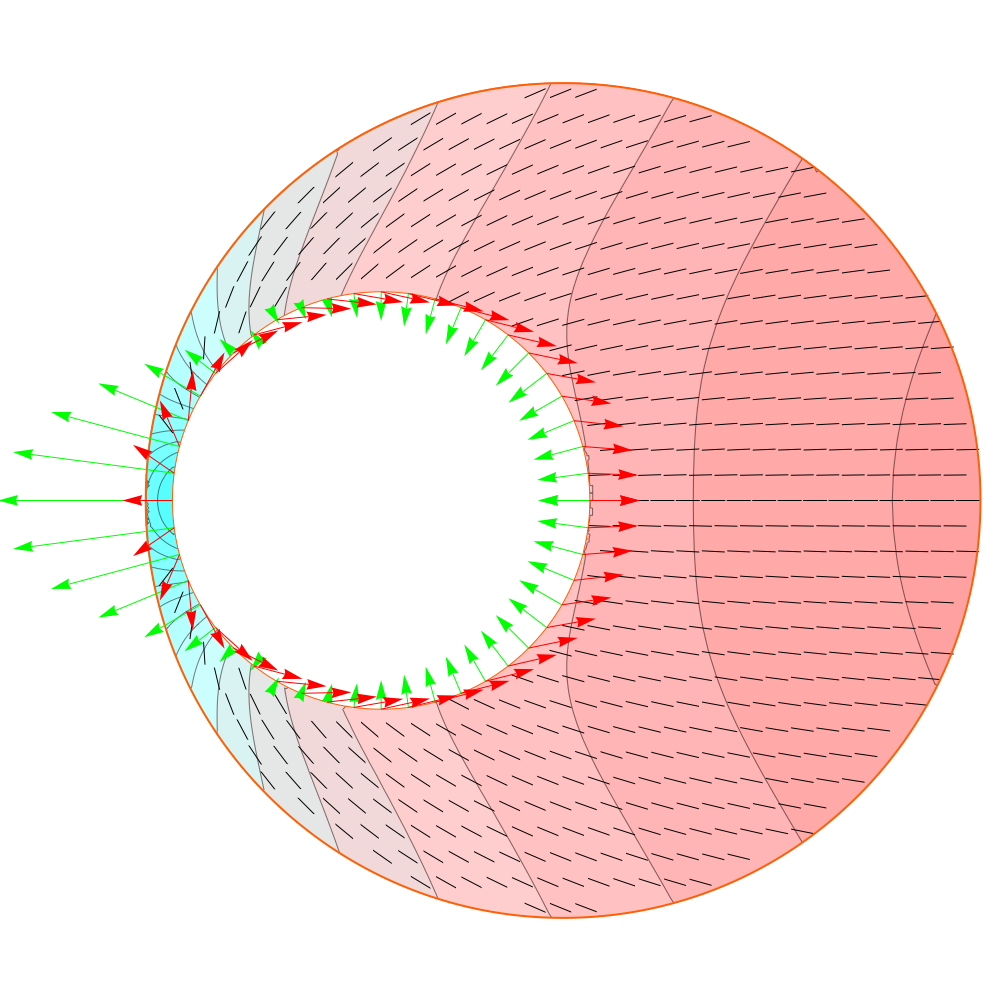}}
  \end{center}
  \caption{(a) Streamfunction and (b) pressure (and nematic
    orientation) visualization for $\alpha = 4.8$ for the sharply
    aligned model at its $r_o = 0.872$ equilibrium point.  The contour levels for the weak flow in
    (a) are within $\psi \pm 0.009$, which are modestly larger than in
    figure~\ref{fig:viz-fixed} (a), and (b) $p \pm 7.42$, which matches
    figure~\ref{fig:viz-fixed} (b).  The arrows also match the definitions
    of figure~\ref{fig:viz-fixed}.}\label{fig:viz-fixed-nematic}
\end{figure}

\begin{figure}
  \begin{center}
  \begin{tikzpicture}
    \begin{axis}
      [ 
      xmin = 0.83,
      xmax = 0.92,
      ymin = -4.1,
      ymax = 5,
      xlabel={$r_o$},
      ylabel={$ \int_\circ \ero \! \cdot \bbf \, ds$},
      tick scale binop=\times,
      width=0.65\textwidth,
      height=0.4\textwidth,
      grid=major, 
      max space between ticks=25,
      legend style={	at={(axis cs:0.84,3.9)},anchor=west,legend
        columns=2,draw=none,legend cell align=left},     
      ]
      \addplot[samples=10, domain=0.8:1, color=gray, forget plot] (x,0) \closedcycle;
      \addplot[mark=*, green!50!black, thick, mark options={fill=white,draw=green!50!black}] table[x expr=\thisrowno{0}, y expr=\thisrowno{2}, col sep=space]{Figures/netforces-conformal.dat};
      \addplot[mark=*, green, thick, mark options={fill=white,draw=green}] table[x expr=\thisrowno{0}, y expr=\thisrowno{1}, col sep=space]{Figures/netforces-conformal.dat};
      \addplot[mark=*, red, thick, mark options={fill=white,draw=red}] table[x expr=\thisrowno{0}, y expr=\thisrowno{3}, col sep=space]{Figures/netforces-conformal.dat};
      \addplot[mark=*, blue, thick, mark options={fill=white,draw=blue}] table[x expr=\thisrowno{0}, y expr=\thisrowno{4}, col sep=space]{Figures/netforces-conformal.dat};
      \legend{
        \footnotesize{{\color{green!50!black} $\frac{1}{\alpha}\int_\circ \ero \! \cdot (p \bI- 2 \bE)\cdot\bn\,ds$}} \;\;\;,
        \footnotesize{{\color{green} $\frac{1}{\alpha}\int_\circ\ero \! \cdot p\bn\,ds$}},
        \footnotesize{{\color{red} $\frac{1}{\alpha}\int_\circ \alpha\,\ero \!\cdot\bD\cdot\bn \,ds$}}\;\;\;,
        \footnotesize{{\color{blue} $\int_\circ \ero\! \cdot\bbf_{\text{total}}\,ds$}},
      }
    \end{axis}
  \end{tikzpicture}
  \end{center}
  \caption{Wall-ward force contributions on the free-floating circle ($\ero = \bx_o/r_o$) for the sharply-aligned model for $\alpha = 4.8$.  Note that for plotting clarity the total force $\bbf_{\text{total}} = (p \bI - 2 \bE)\cdot\bn-\alpha \bD\cdot\bn$ (blue) is exaggerated by an $\alpha$ factor with respect to the others to better show the $r_o \approx 0.872$ equilibrium radius.}\label{fig:netalignedforces}
\end{figure}
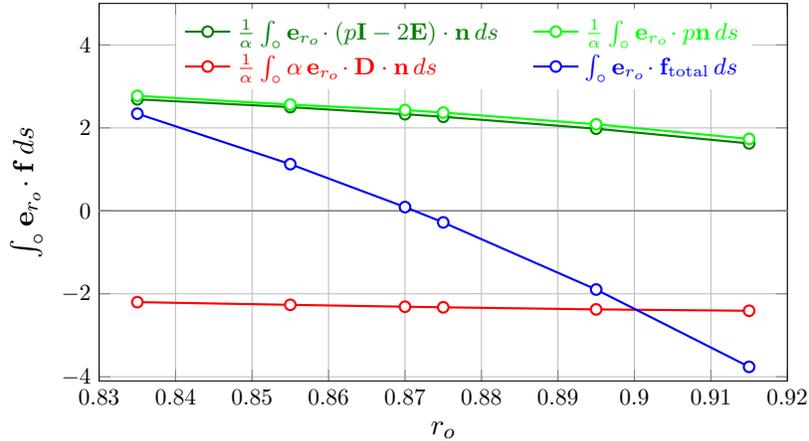

Comparing figure~\ref{fig:viz-fixed} with figure~\ref{fig:viz-fixed-nematic} shows that the nematic direction for this model and the corresponding simulation result are in close agreement.
The implied $\bD$ field can be used to estimate an effective distributed force $\nabla\cdot \alpha\bD$ for a viscous flow in the same geometry.  The response to this force is primarily hydrostatic, and pressure visualized in figure~\ref{fig:viz-fixed-nematic} (b) compares well with the pressure from the full simulation in figure~\ref{fig:viz-fixed} (b).  The net $\alpha\bD$ surface traction, which primarily draws the object to the center of the container, is countered almost entirely by the pressure.  The net force contributions on the object for different locations are shown in figure~\ref{fig:netalignedforces}.  A stable equilibrium of $\req = 0.872$ is predicted, which is close to the value for the full simulation ($r_o^{eq} = 0.855$). We also see in figure~\ref{fig:fixed-prms} (b) that the full simulation has $\req \to 0.90$ for $\zeta \to \infty$, which is only slightly larger than the  $\req = 0.872$ prediction for the sharply aligned model, which is remarkable agreement considering that the model neglects all active suspension dynamics and surface tractions aside from $\bn \cdot \alpha \bD$.   At all points, the predicted $\alpha \bD$-driven Newtonian flow viscous contribution is small.  Both the pressure attraction toward the wall and net active tension $\alpha\bD$ away from it decrease with distance, and both are significantly stronger than their difference.   This indicates that although the flow is critical for bringing the object to the fixed-point, it is not a primary participant in the force balanced reflected in this configuration.  The principal difference in the flow between the full simulation in figure~\ref{fig:viz-fixed} (a) and the sharply-aligned model in figure~\ref{fig:viz-fixed-nematic} (a) is the secondary circulation pair to the right of the object in the full simulations.   This feature is seen in figure~\ref{fig:viz-approach-fixed} to only appear slowly, after the object is essentially fixed.  The cause of the low pressure in the gap is simply the gap-parallel component of the active contraction of the aligned agents pulling on the incompressible fluid in the narrow gap.

The corresponding defect-free arrangement with $m_\circ=+1$ and circumferential rather than radial nematic structure was briefly introduced as an initial condition in the previous section, although, as mentioned, such an arrangement is never observed to persist.  We can examine this in the sharply aligned limit by taking $\theta(\phi) = \phi + \smash{\frac{\pi}{2}}$ in (\ref{e:wsol}) and applying the same conformal mapping.  The result is a change in the signs of all tractions on the object.  The active stress $\alpha\bD$ also changes sign, which in turn changes the signs of the flow velocity and pressure.  Overall, the net forces in figure~\ref{fig:netalignedforces} all switch sign.  With this switch the $r_o\approx 0.872$ equilibrium point remains, but it is unstable, consistent with its absense from the full simulations.

\section{Limit-cycle Behavior}
\label{s:limit-cycle}

\subsection{Flow characteristics}

\begin{figure}
  \begin{center}
    \subfigure[$t=0$]{l\includegraphics[width=0.32\textwidth]{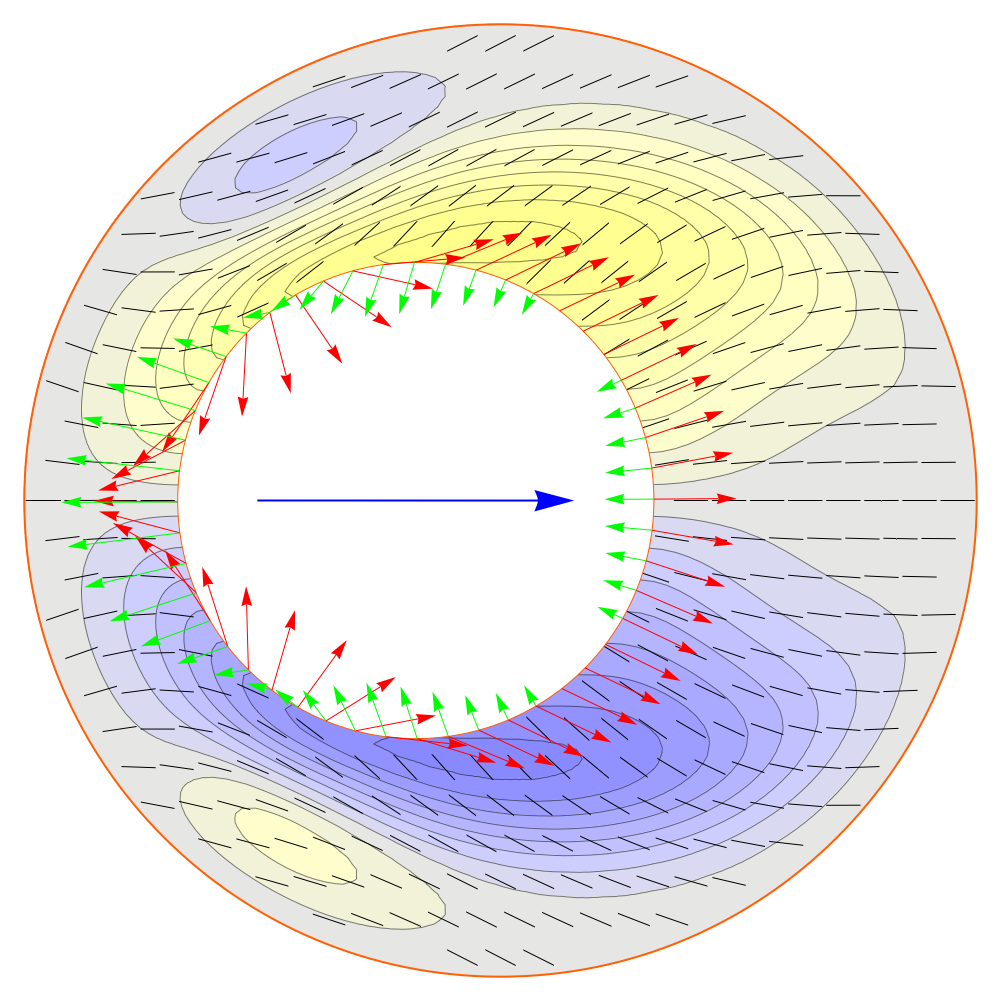}}
    \subfigure[$t=10$]{\includegraphics[width=0.32\textwidth]{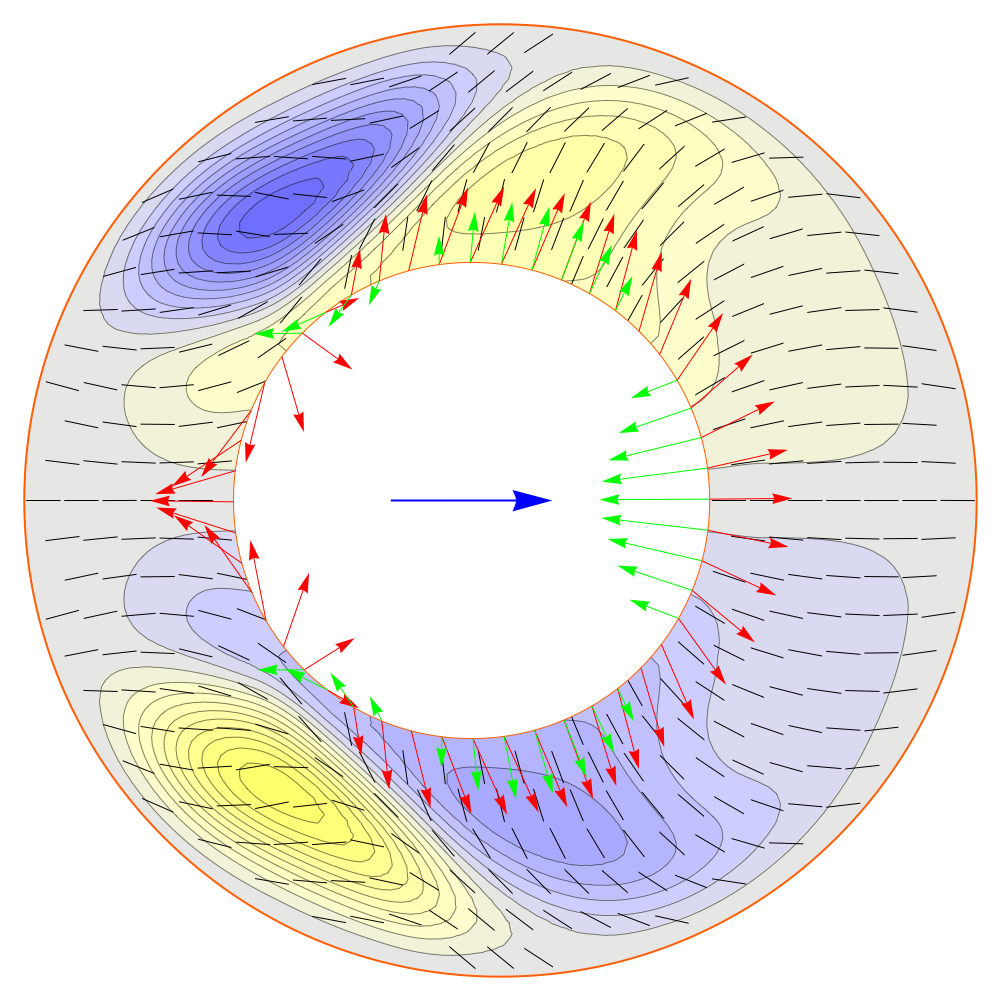}}
    \subfigure[$t=15$]{\includegraphics[width=0.32\textwidth]{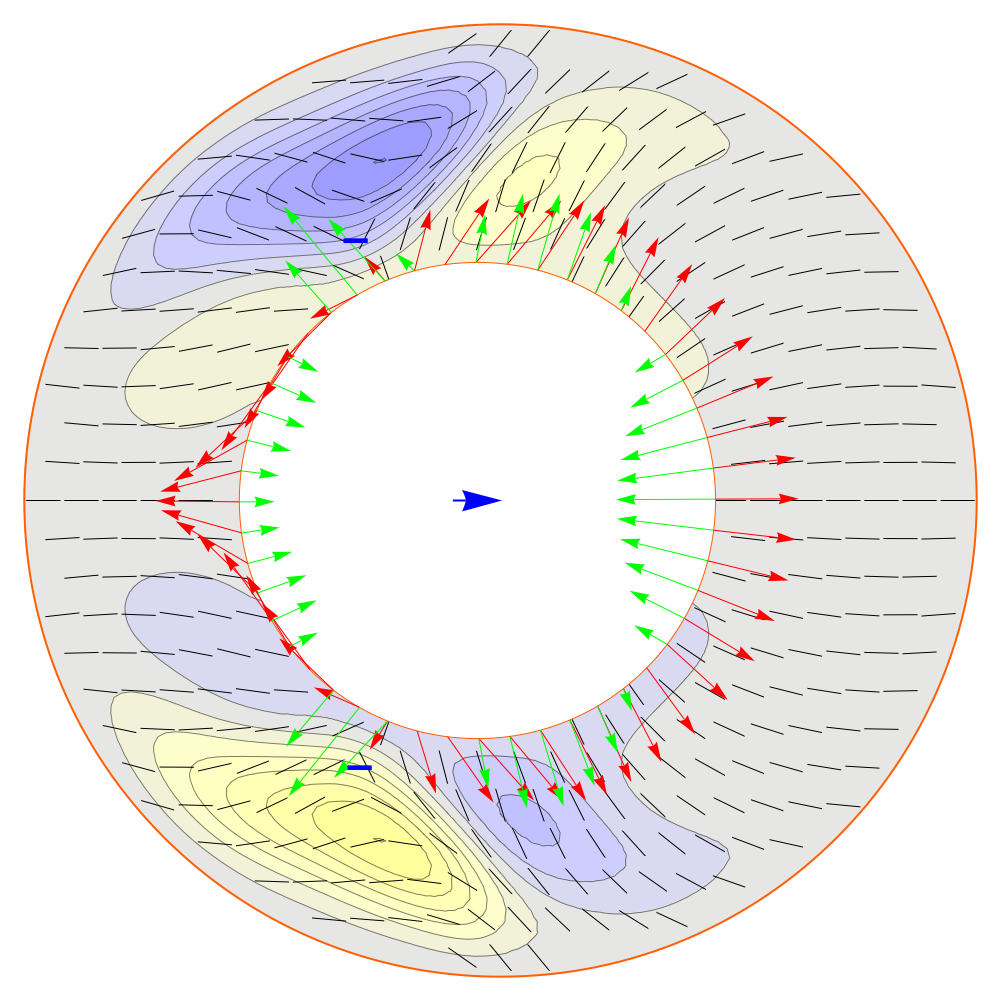}}
    \subfigure[$t=30$]{\includegraphics[width=0.32\textwidth]{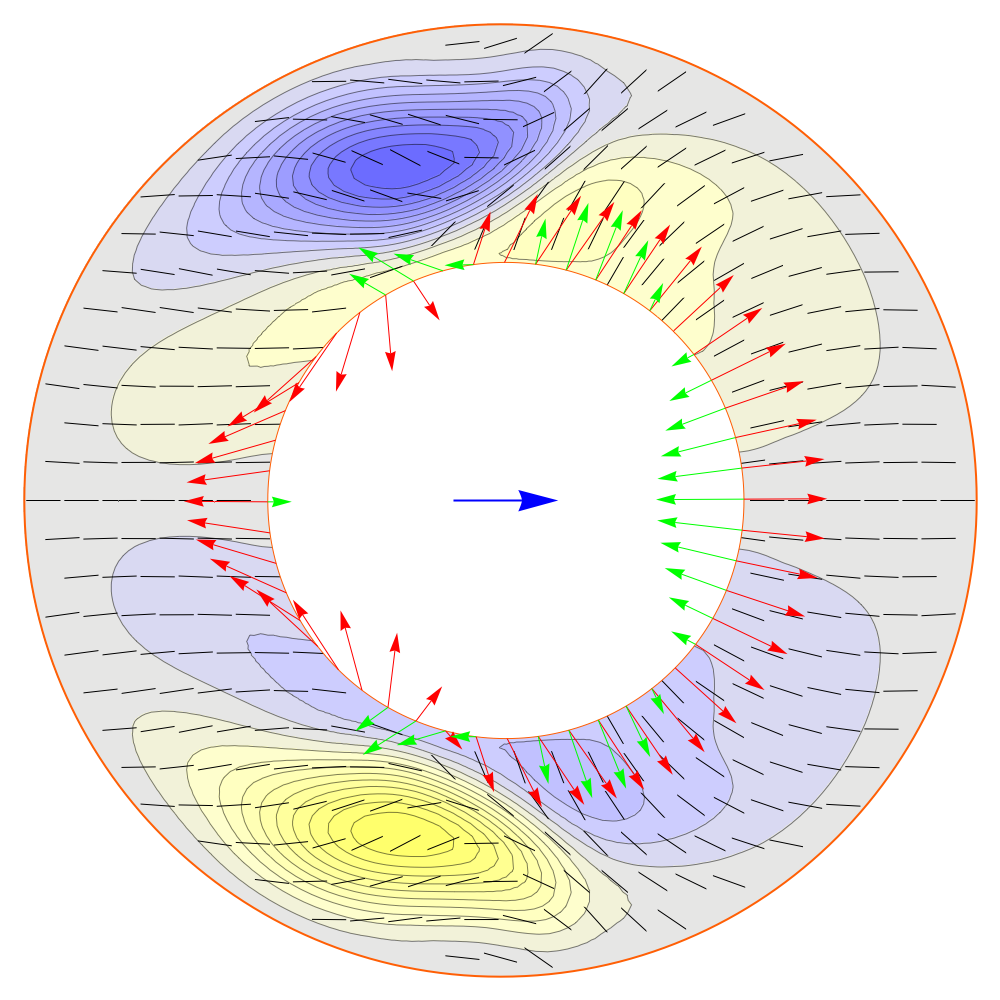}}
    \subfigure[$t=50$]{\includegraphics[width=0.32\textwidth]{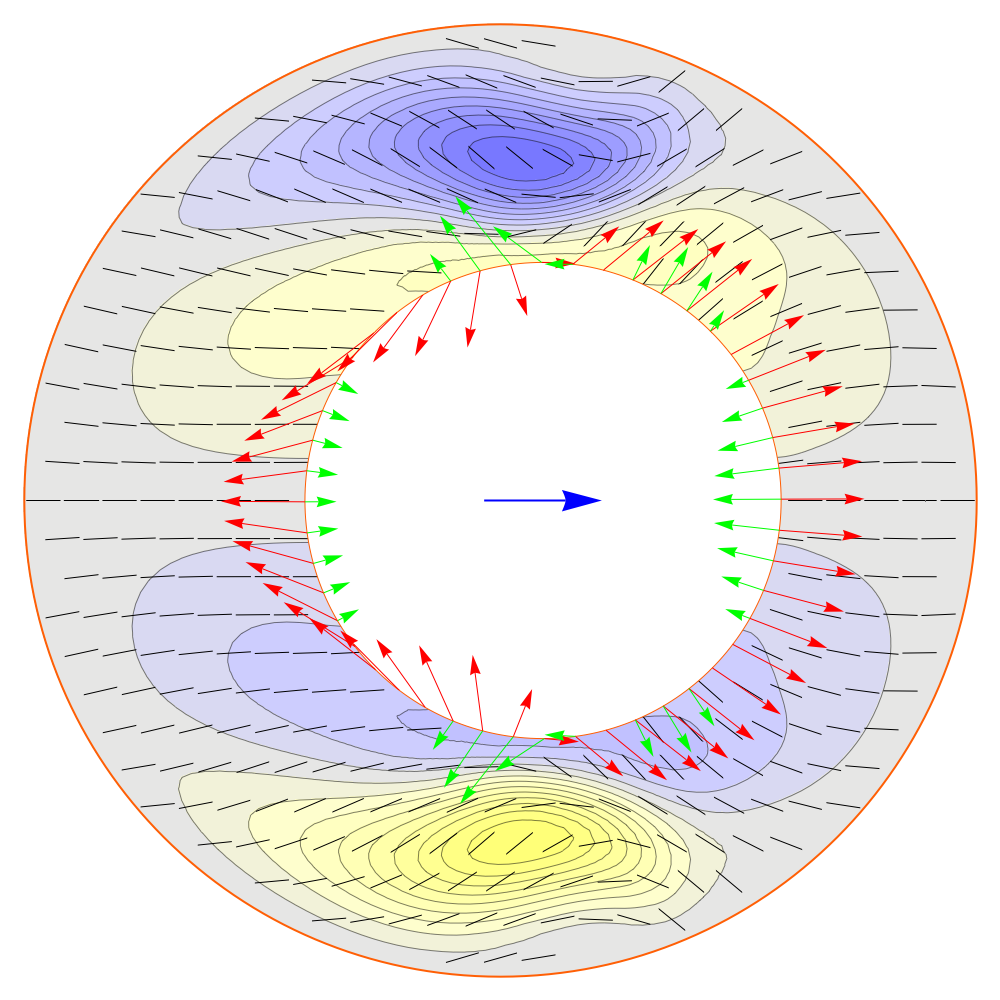}}
    \subfigure[$t=90$]{\includegraphics[width=0.32\textwidth]{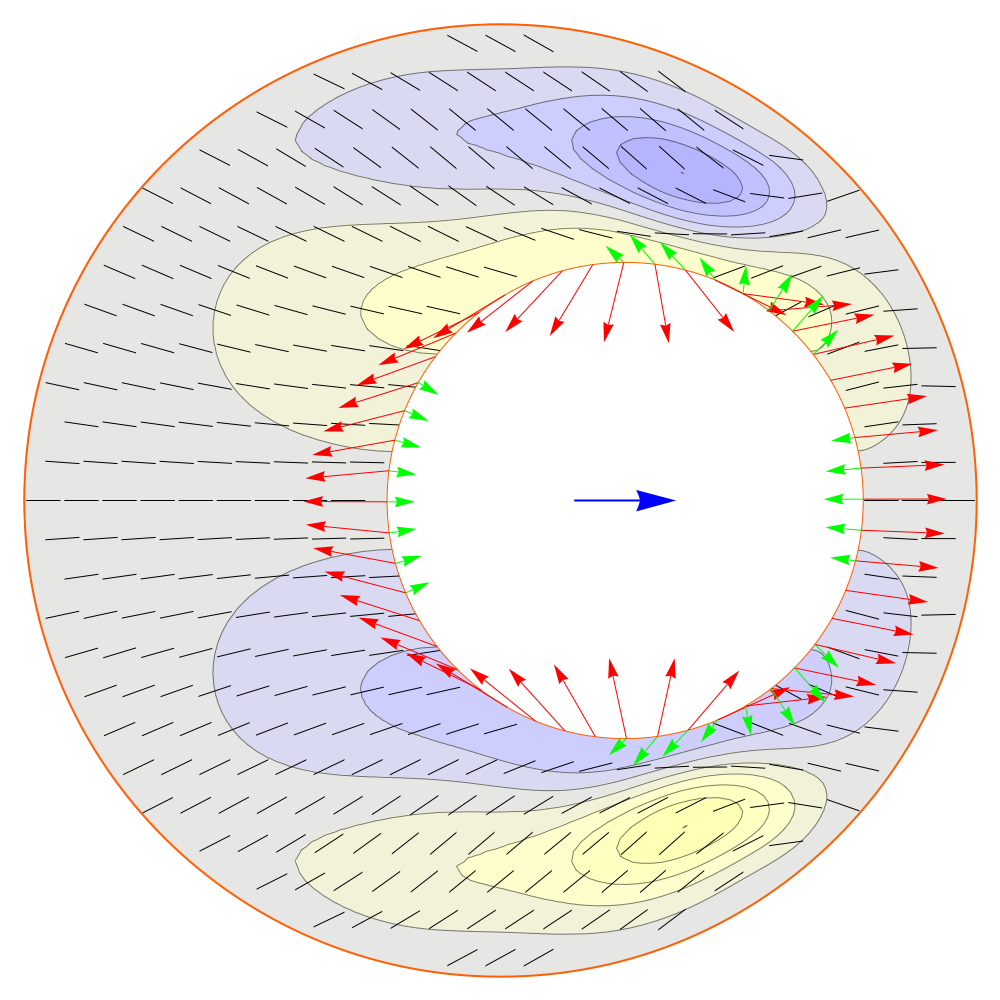}}
    \subfigure[$t=100$]{\includegraphics[width=0.32\textwidth]{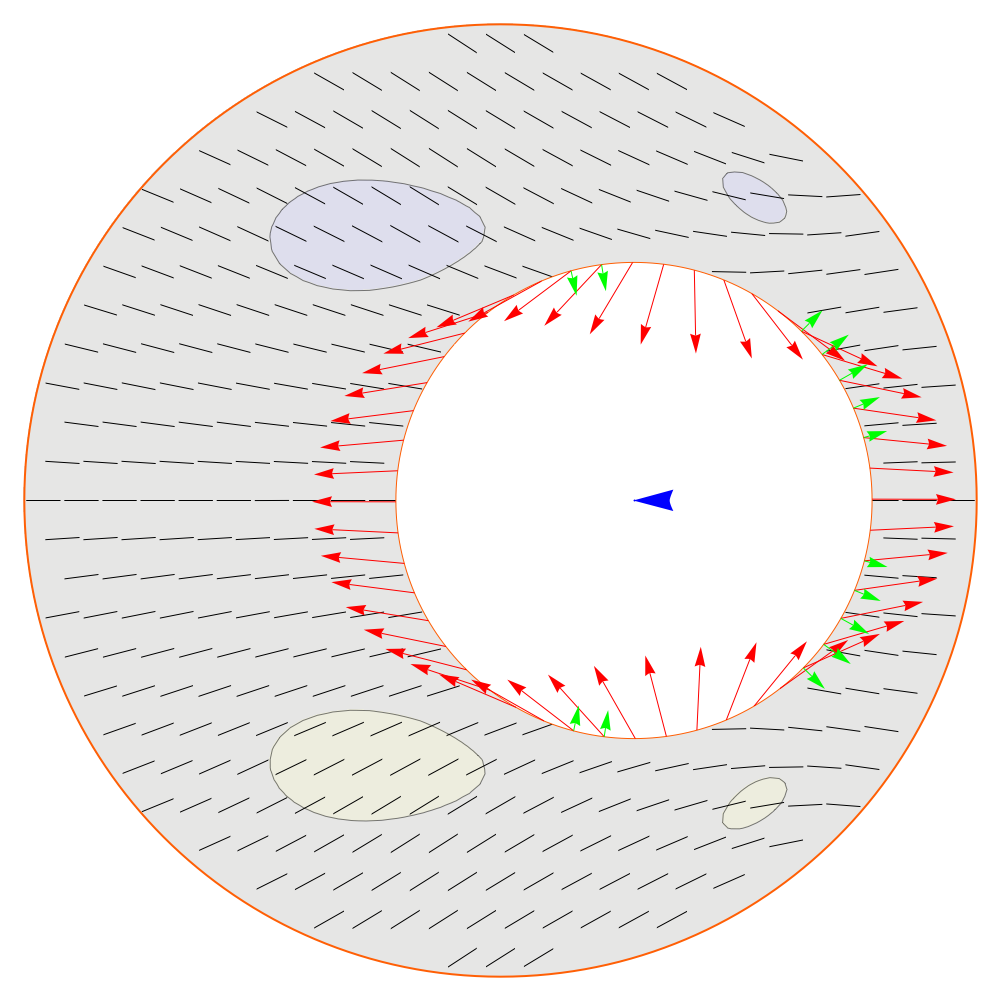}}
    \subfigure[$t=110$]{\includegraphics[width=0.32\textwidth]{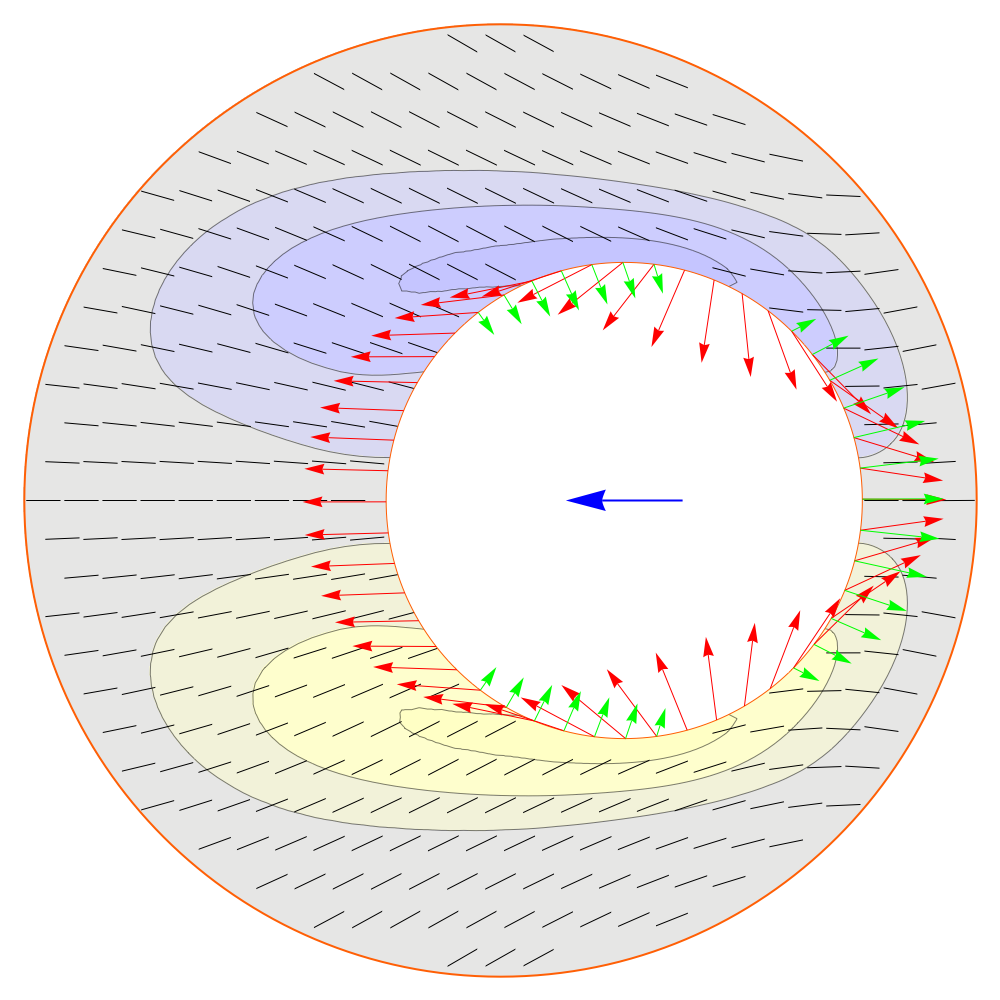}}
    \subfigure[$t=120$]{\includegraphics[width=0.32\textwidth]{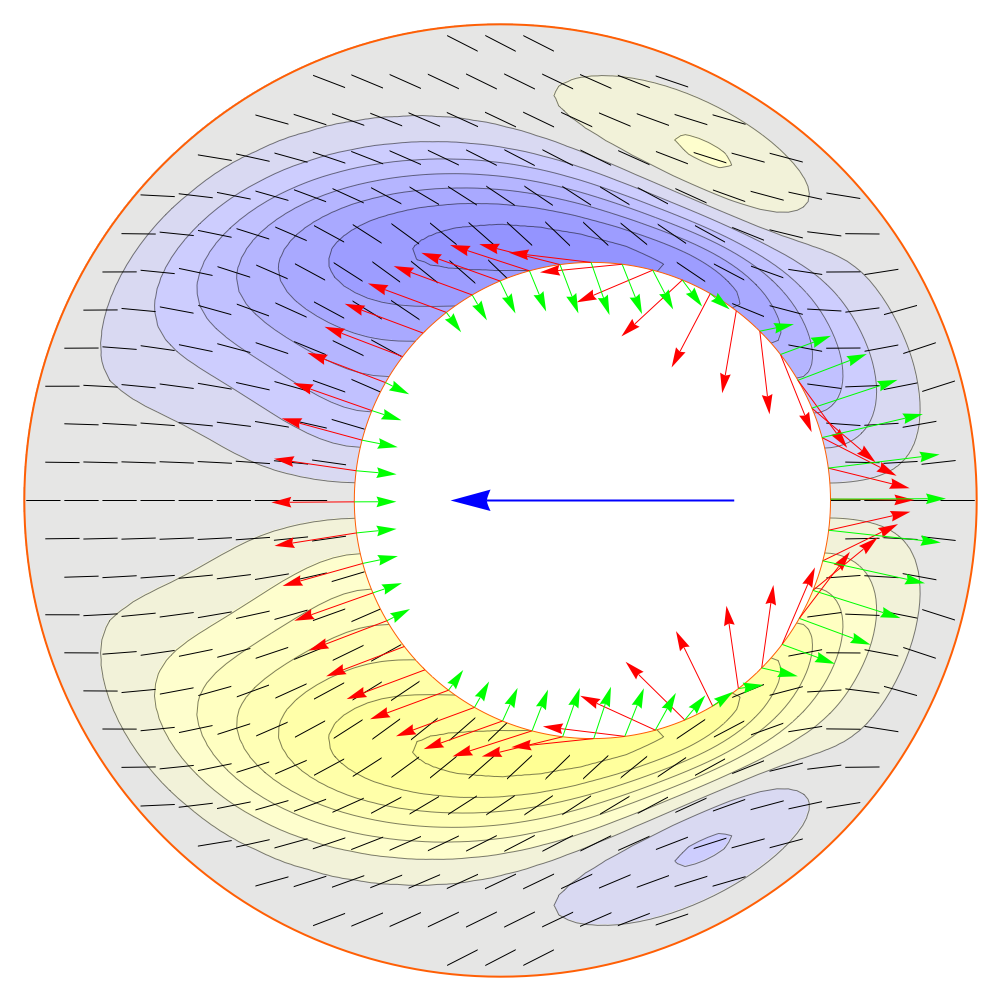}}
  \end{center}
  \caption{Streamfunction and nematic orientation visualization for times labeled for $\alpha = 4.8$, $\zeta = 1.0$ in a case that evolves into a limit cycle.  The streamfunction levels are equally spaced between $\psi = \pm 0.031$, the peak values for this time series.  The domain is rotated so the object is moving horizontally; the actual orientation is shown in  figure~\ref{fig:basictraj} (b).  Note that the times shown as labeled are hand selected to illustrate changes.  The smaller arrows visualize the surface tractions:  red is the active deviatoric traction $\alpha (\bD-\bI/2) \cdot\bn$ term and green is the net hydrodynamics traction $(-p\bI + 2 \bE)\cdot\bn$.  The larger blue arrow has length proportional to the velocity of the object.  The $-$ symbols in (c) indicate the location of the $m=-\shalf$ nematic defects that appear only around this time. }\label{fig:streamviz-limit}
\end{figure}

A half-period of the cycle for the case of figures~\ref{fig:basictraj} (a) and (b) is visualized in figure~\ref{fig:streamviz-limit}.   The basic arrangement has a uniform and defect-free nematic field  ($\theta \approx 0$ as visualized) with $m_\circ = 0$.  The only significant surface traction is the $\alpha \bD$ contribution, which is almost entirely balanced by the hydrodynamics stresses.  Were they included in the visualizations, the corresponding tractions for the other $\beta \bS\cddot \bE + \beta \zeta (\bD\cdot\bD - \bS\cddot \bD)$ components would have arrows well smaller than the smallest surface traction arrows visualized in figure, even if multiplied by a factor of 10.

\begin{figure}
  \begin{center}
    \subfigure[$t=5$]{\begin{tikzpicture}
      \node[anchor=south west, inner sep=0] (image) at (0,0)  
      {\includegraphics[width=0.32\textwidth]{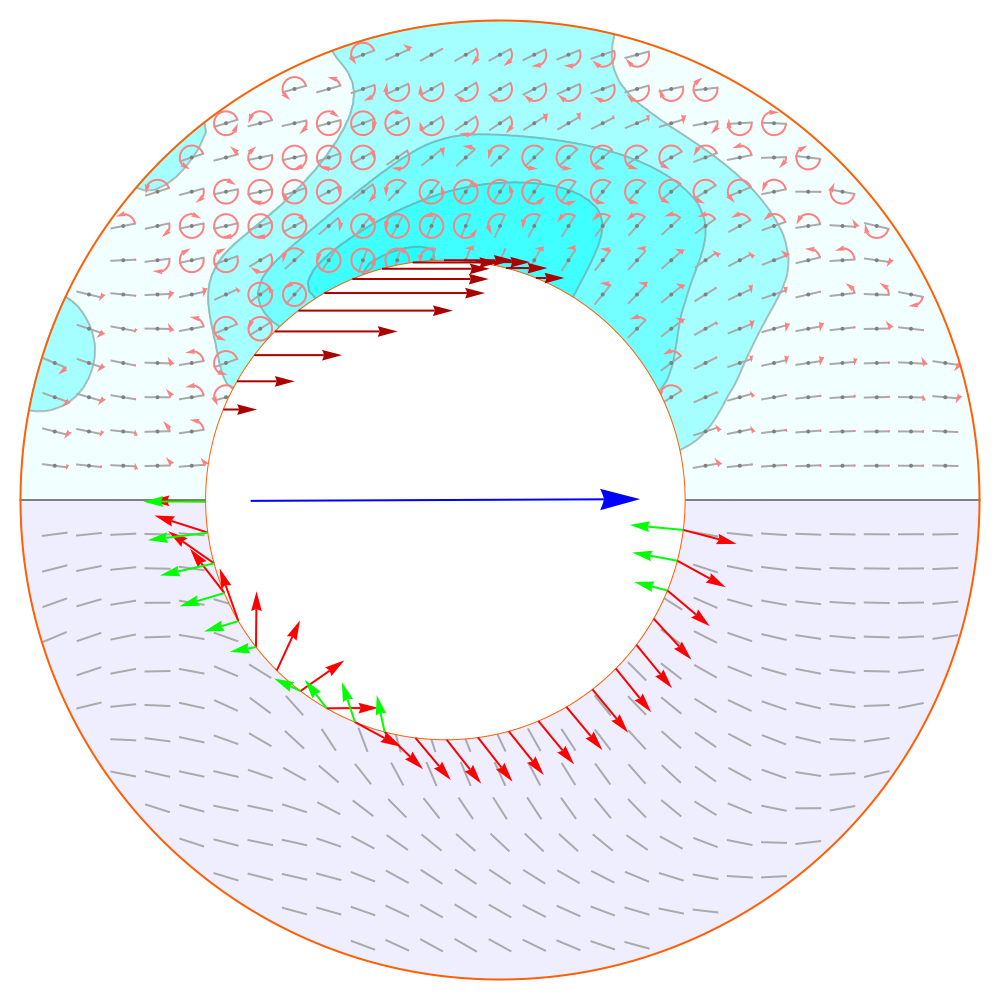}};
      \begin{scope}[x={(image.south east)},y={(image.north west)}]
        \node at (0.10,0.550){ {\color{cyan} $|\theta|$}};
        \node at (0.1,0.45){ {\color{blue} $\lambda_1$} };
        \node at (0.46,0.56) { \begin{minipage}{0.12\textwidth}\tiny\color{red!60!black}$\alpha[\be_x \cdot (\bD -\bI/2) \cdot \bn$\\\hspace*{0.1in} $- D_o \cos \phi]\be_x$\end{minipage}};
        \node at (0.48,0.380){ \tiny {\color{red} $\alpha \bD  \cdot \bn$} };
        \node at (0.47,0.45){\tiny {\color{green} $(2 \bE - p\bI)\cdot \bn$}};

      \end{scope}
    \end{tikzpicture}}
    \subfigure[$t=10$]{\includegraphics[width=0.32\textwidth]{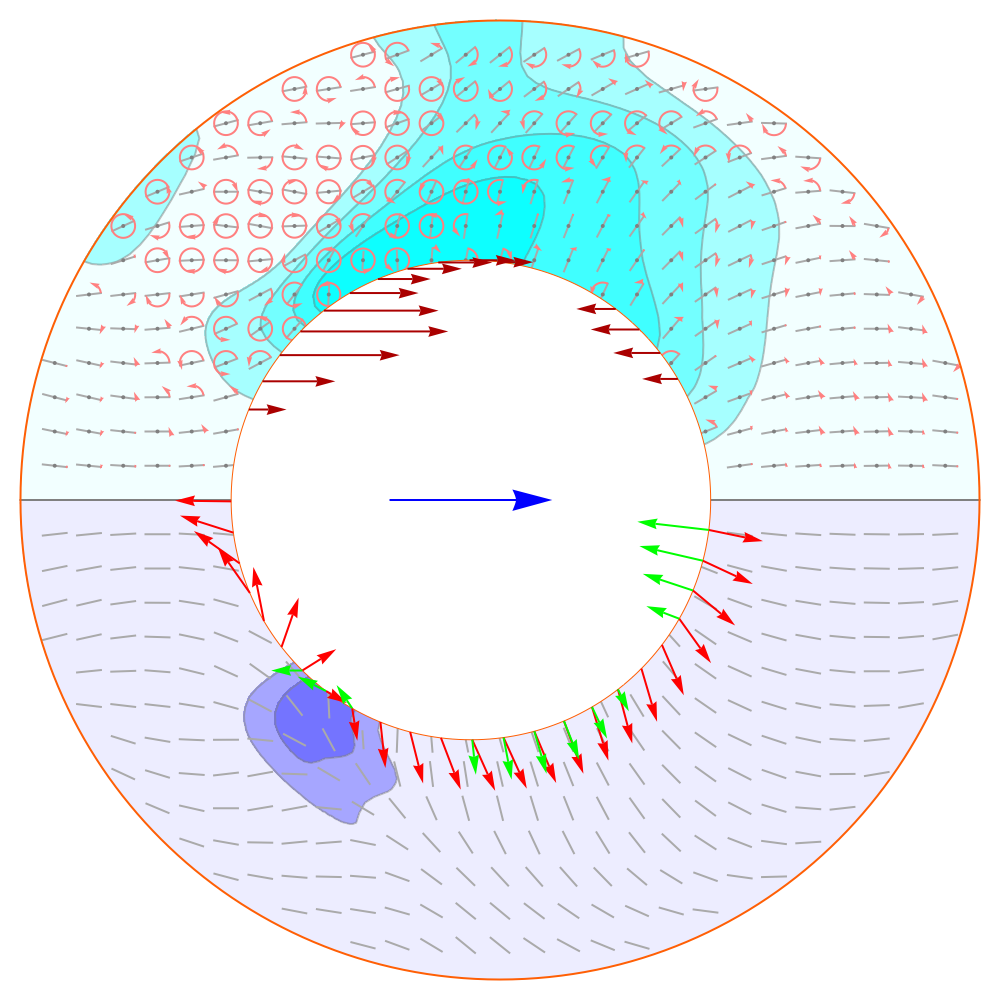}}
    \subfigure[$t=15$]{\includegraphics[width=0.32\textwidth]{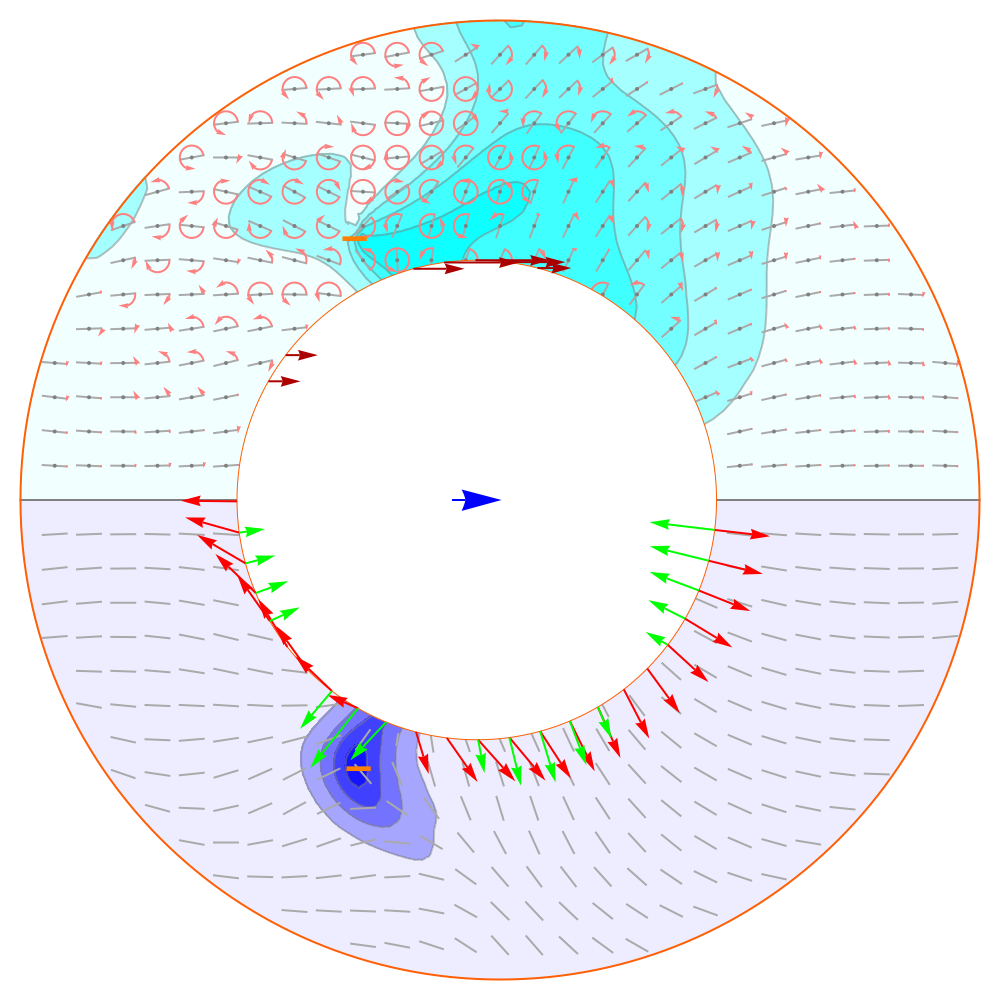}}\\
    \subfigure[$t=70$]{\includegraphics[width=0.32\textwidth]{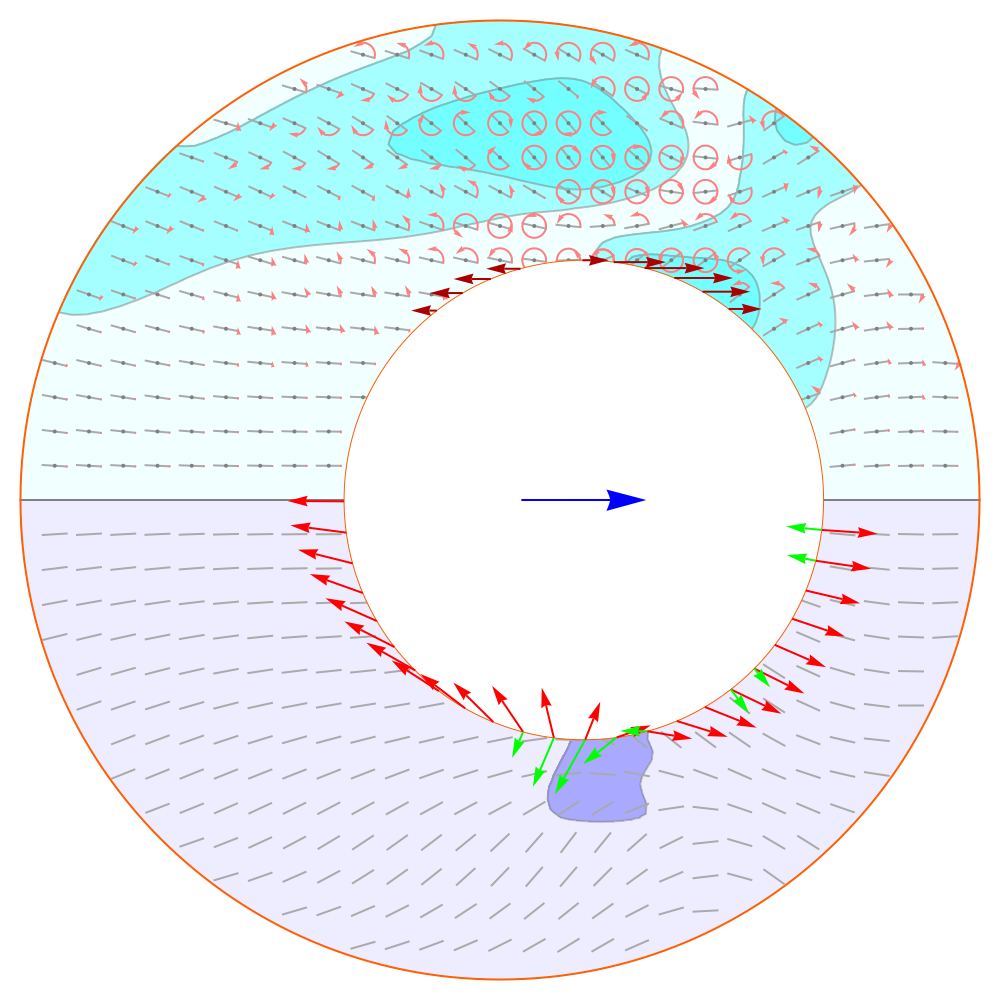}}
    \subfigure[$t=90$]{\includegraphics[width=0.32\textwidth]{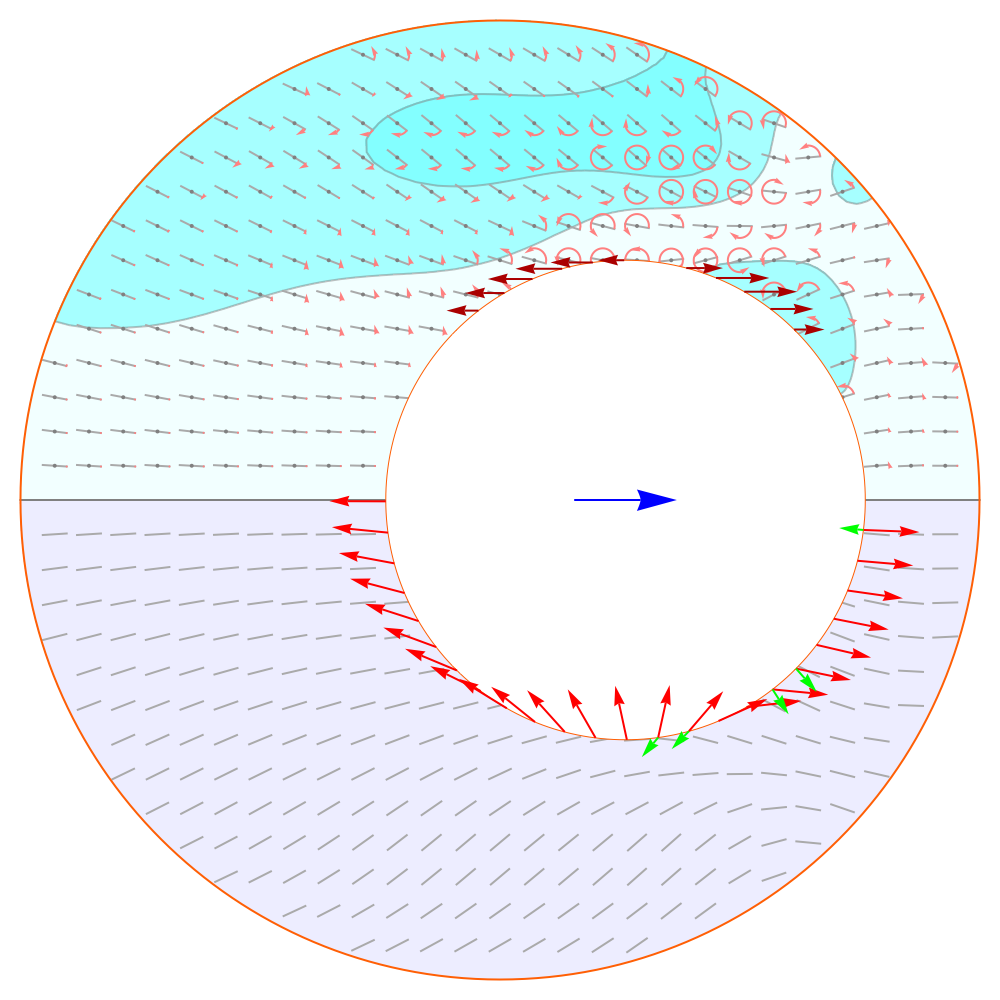}}
    \subfigure[$t=100$]{\includegraphics[width=0.32\textwidth]{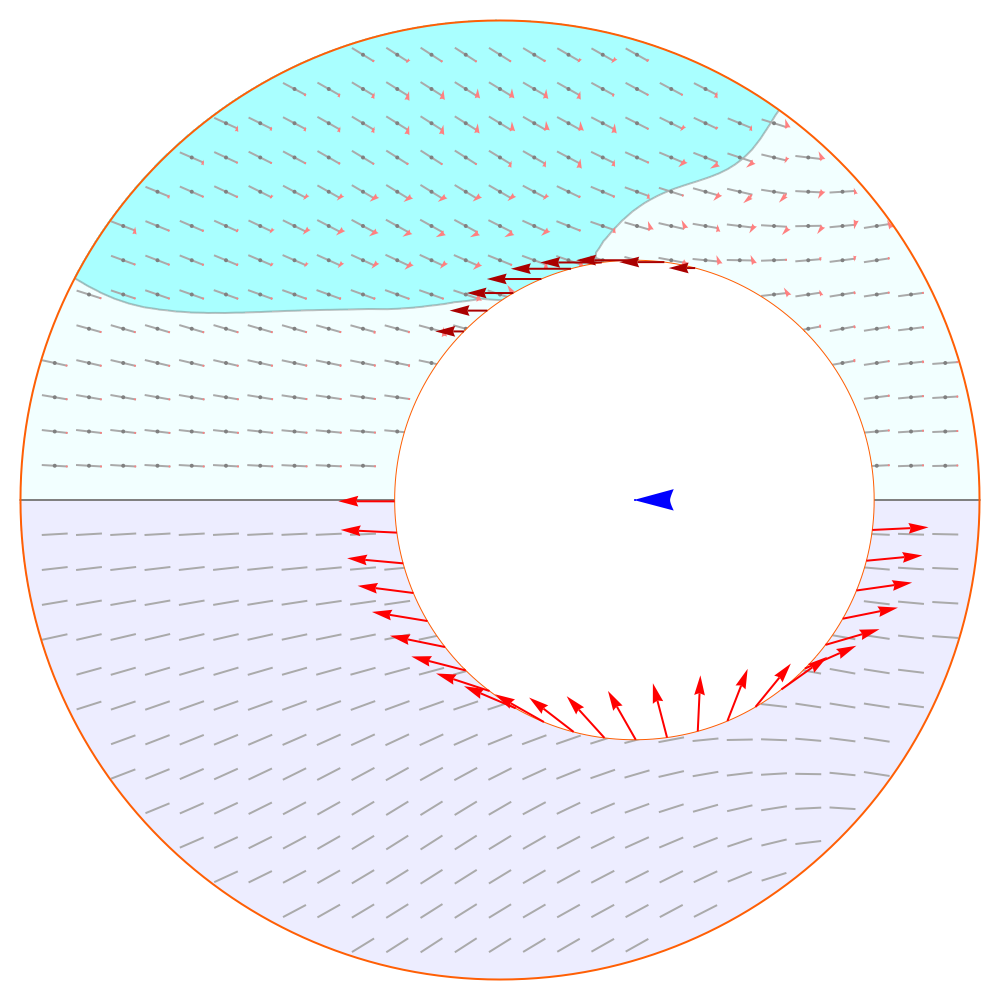}}
  \end{center}
  \caption{Visualization of key parts of the limit-cycle behavior for the
    times labeled with respect to frame (a):  (a--c)
    rapid deceleration and (d--e) reversal.  In the top half of each, cyan levels indicate local nematic deflection $|\theta|$, with color from horizontal (light) to vertical (darker) in intervals of $\pi/10$, and the surface arrows are proportional in length to the horizontal deviatoric $\alpha (\bD-\bI/2)$ component minus its at-rest value for a $m_\circ=0$ defect-free base state: $\be_x \cdot \alpha (\bD -\bI/2) \cdot \bn - 0.983 \alpha \cos \phi$, where $\be_x = (1,0)$.  The blue levels in the bottom half indicate $\lambda_1$ with indigo level at intervals of 0.1 between 0 (lighter) and 0.5 (darker), and the surface arrows are proportional to $\alpha (\bD-\bI/2) \cdot\bn$ (red) and $(- p\bI + 2 \bE)\cdot\bn$, which are scaled to 1/2 the length of those in the top half of each frame.  In the top half, the orange arcs associated with the segments indicate the sense and amplitude of the local vorticity, providing a measure of the flow kinematic rotation.}\label{fig:limit-ang-l}
\end{figure}
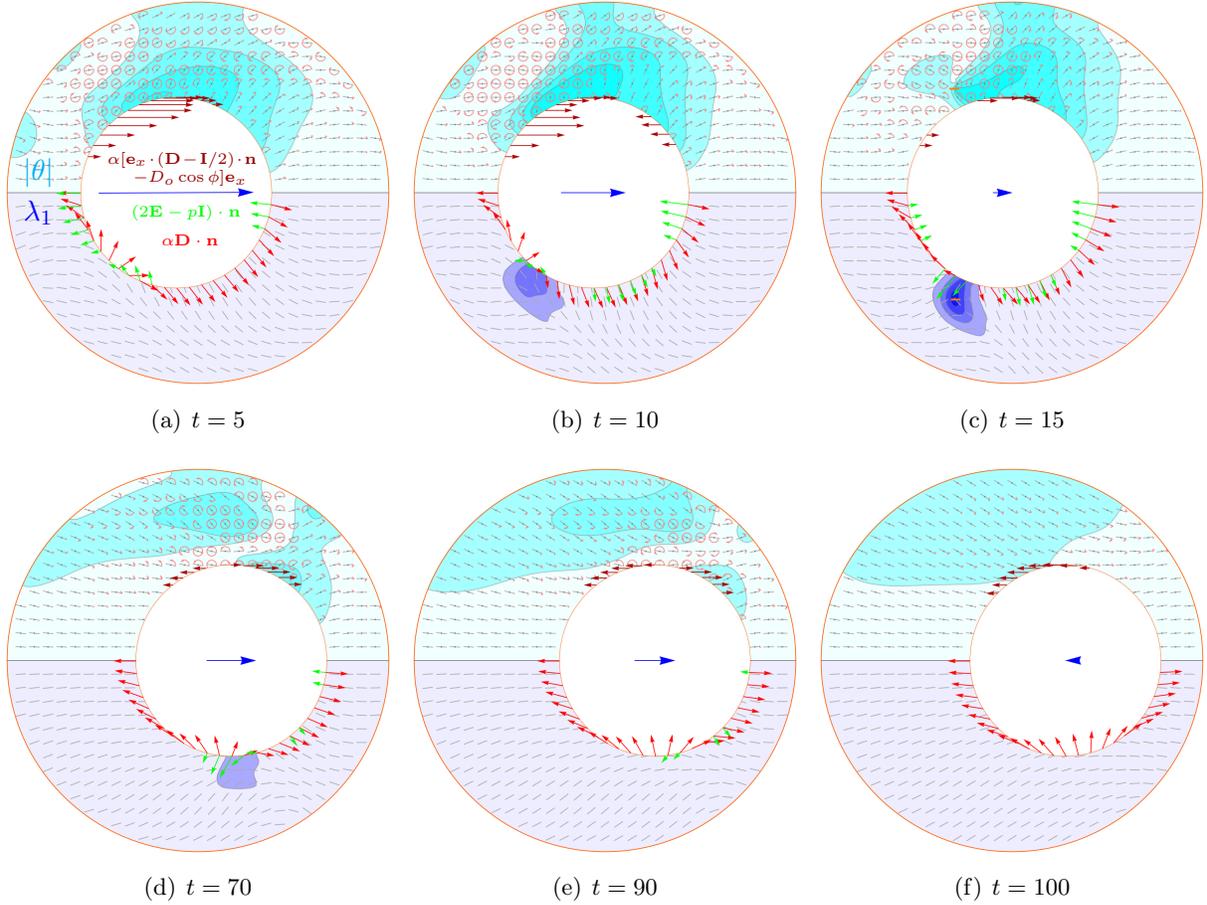

Figure~\ref{fig:streamviz-limit} (a) shows the flow at the highest object speed, which is reached early in its crossing of the container (defined here to be at $t=0$).  Without flow, the contractors would pull on both the front and rear of the object.  The component in the direction of travel is $f_x = \alpha D_o \cos\phi$, where $\phi = 0$ on the $x$-axis.  This stage of the motion seems best explained as a disruption of this balance as is shown in figure~\ref{fig:limit-ang-l} (a)--(c) with vectors $\alpha [\be_x \cdot \bD\cdot \bn - D_o \cos \phi]\be_x$ for $\be_x = (1,0)$.  This figure also shows flow-driven bending of the nematic direction in the wake region.  The resulting nematic reorientation disrupts the force balance of a uniform $\theta =0$ nematic.  For the larger $|\theta|$, there is also a significant off-axis component that applies a horizontal force, augmenting motion.  The time period of acceleration is $t\approx 30$, which is between the $\tau_{u_f}^\ell = 10$ and $\tau_{u_\circ}^\ell = 100$ time scales in table~\ref{tab:taus}, consistent with advection distorting the nematic orientation faster than the container-scale diffusion $t_T^\ell = 100$ that would restore it, leading to this force imbalance.

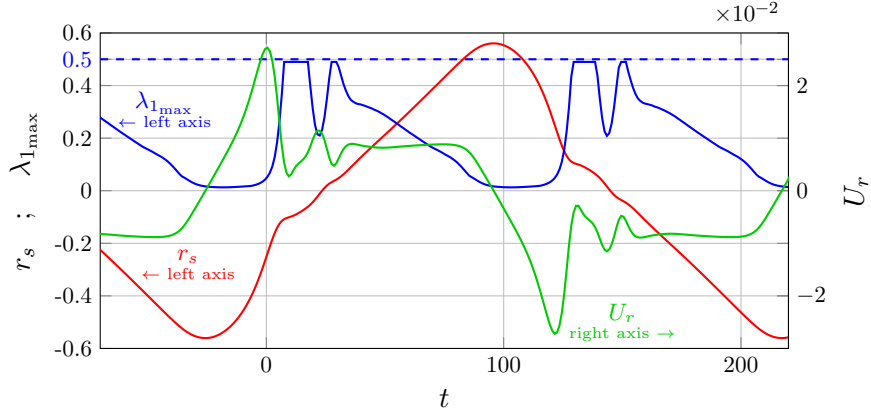
\begin{figure}
  \begin{center}
    \begin{tikzpicture}
      \begin{axis}
        [ 
        ymin = -0.6,
        ymax = 0.6,
        xmin = -70,
        xmax = 220,
        ylabel={$r_s$\;\; ; \;\; $\lambda_{1_\text{max}}$}, 
        xlabel={$t$},
        tick scale binop=\times,
        width=0.65\textwidth,
        height=0.35\textwidth,
        grid=major,
        max space between ticks=100,
        ytick={-0.6,-0.4,-0.2,0,0.2,0.4,0.5,0.6},
        yticklabels={-0.6,-0.4,-0.2,0,0.2,0.4,{\color{blue}0.5},0.6},
        ]
        \addplot+[no marks, thick, color=red] table[x expr=\thisrowno{0}, y expr=\thisrowno{1}, col sep=space] {Figures/url-hist-shift.dat};
        \addplot+[no marks, thick, color=blue] table[x expr=\thisrowno{0}, y expr=\thisrowno{4}, col sep=space]
        {Figures/url-hist-shift.dat};
        \node at (axis cs:-43,0.31) 
        {{\begin{minipage}{0.2\textwidth}\centering \footnotesize \color{blue}
              $\lambda_{1_\text{max}}$ \\ \tiny $\leftarrow$ left axis\end{minipage}}};
       \node at (axis cs:-33,-0.28)
        {{\begin{minipage}{0.2\textwidth}\centering \footnotesize \color{red}
              $r_s$ \\\tiny $\leftarrow$ left axis\end{minipage}}};
        \addplot[samples=10, domain=-70:300, color=blue, dashed, thick, forget plot] (x,0.5);
      \end{axis}
      \begin{axis}
        [
        axis y line*=right,
        axis x line = none,
        yticklabel pos=right,
        ymin = -0.03,
        ymax = 0.03,
        xmin = -70,
        xmax = 220,
        ylabel={$U_r$}, 
        xlabel={$d_T$},
        tick scale binop=\times,
        width=0.65\textwidth,
        height=0.35\textwidth,
        grid=none, 
        max space between ticks=25,
        ]
        \addplot+[no marks, thick, color=green!80!black] table[x
        expr=\thisrowno{0}, y expr=\thisrowno{2}, col sep=space]
        {Figures/url-hist-shift.dat};
        \node at (axis cs:150,-0.025) 
        {{\begin{minipage}{0.2\textwidth}\centering
              \footnotesize\color{green!80!black} $U_r$ \\ \tiny  right axis $\rightarrow$\end{minipage}}};
      \end{axis}
    \end{tikzpicture}
  \caption{The transient signed object radial velocities ($U_r= |\bU| \sgn U_x$) and signed radial position ($r_s = r_o \sgn x$) and defect formation quantified by $\max \lambda_1$ for $\alpha = 5$, $d_T = 0.01$ and $\zeta = 0.7$. }\label{fig:hists}
  \end{center}
\end{figure}

As the object reaches this peak velocity, the nematic orientation changes most significantly near $\phi = 3\pi/4$ on the object, as seen in figure~\ref{fig:limit-ang-l} (a) with closely-spaced $|\theta|$ contours in this region.  Starting at $t\approx 10$, the nematic structure fractures (figure~\ref{fig:limit-ang-l} b) with appearance of $-\shalf$ disclinations just off the object by $t=15$ (figure~\ref{fig:streamviz-limit} c and figure~\ref{fig:limit-ang-l} b), which together leave the object temporarily with $m_\circ = +1$. After this phase, prominent counter-rotating circulations appear across the top and bottom of the container, with peak velocities several times that of the object.  This deceleration is seen in figure~\ref{fig:hists} to correspond with a jump of $\lambda_1$ from near 0 to nearly 0.5, consistent with the formation of the diffusion-regularized defects.  These changes are rapid, aligning with the $\tau_{u_f}^\delta = \tau_T^\delta = \tau_\zeta = 1$ time scales of table~\ref{tab:taus}.

The defects and associated crack-like features in the nematic field do two things.  Nearby, where $\lambda_{1_\text{max}} \to 0.5$ indicates local isotropy, the $\alpha \bD$ term is balanced only by pressure, not flow inducing.  This decreases the object pushing component, though it still fails to balance the pulling component on the leading side.  However, more consequentially, the nominal crack in the nematic structure allows the fluid behind the object to rapidly relax on the fast $\tau_\zeta=\tau_T^\delta=1$ time scales to $\theta \approx 0$ nematic orientation, so it again balances the corresponding pulling component on the leading side, as it would for uniform $\theta=0$.  As a consequence, object slows significantly, though it continues to be drawn forward slowly by the horizontal component of the $\theta < \pi/2$ oriented nematic ahead of the defects and cracks.  The slowed motion allows strain to relax and the crack to close, although for some cases it transiently reopens as the object again accelerates, repeating the same series of events, as indicated by the second less pronounced deceleration and increase of $\lambda_1$ in figure~\ref{fig:hists}.

The motion is reversed at the end of a cycle, but more quickly than the observed instabilities of the baseline $\theta = 0$ state.  The particular nematic structure that remains after the recent transit is important in the reversal.  Figure~\ref{fig:basictraj} shows that even once an apparent oscillation starts, it takes several cycles to settle into the fully periodic behavior, suggesting that the oscillation time scale $\tau_{u_o}^\ell$ is linked to the container-scale diffusion $\tau_T^\ell=100$, the only other long time scale in table~\ref{tab:taus}.  The remnants of the previous crossing  are seen to evolve on this slow diffusion time scale in figure~\ref{fig:limit-ang-l} (d--f).  As the object finishes its traverse, its unbalanced surface tractions are associate with two regions where the nematic structure is misaligned with the object trajectory (figure~\ref{fig:limit-ang-l} e).  Considering the $y>0$ (top half) region ahead of the object, there is a small $\theta > 0$ region near $\phi = \pi/4$, and above and extending behind it there is a large  $\theta < 0$ region, most pronounced away from the object.  The small region more quickly adjusts to uniform $\theta=0$ via thermal diffusion $d_T$.  The larger region, by virtue of its scale, decays more slowly, spreading toward the object.  Reversal occurs approximately at the time when the balance between the larger and smaller regions shifts.  This accelerates the object, leading to nematic field rotation and significant acceleration, and eventually the cracks forming again.  This difference in rates of the relaxation of the larger and smaller $|\theta| > 0$ regions is simply due to the quadratic dependence of diffusion time on length scale.

\clearpage

\subsection{Parametric dependencies}

\begin{figure}
  \begin{center}
     \subfigure[]{
    \begin{tikzpicture} 
      \begin{axis}
        [ 
        ymin = 0.00,
        ymax = 0.80,
        xmin = 4,
        xmax = 7.5,
        ylabel={amplitude -- $r_o^{\text{max}}$}, 
        xlabel={$\alpha$},
        tick scale binop=\times,
        width=0.42\textwidth,
        height=0.35\textwidth,
        grid=major,
        max space between ticks=25,
        ]
        \addplot+[no marks, thick, color=red] table[x
        expr=\thisrowno{0}, y expr=\thisrowno{1}, col sep=space] {Figures/alpha-amp-limit.dat};
        \node at (axis cs:6.4, 0.7) {\begin{minipage}{0.25\textwidth}{\footnotesize \color{red} amplitude $r_o^{\text{max}}$\\  left axis $\leftarrow$}\end{minipage}};
      \end{axis}
      \begin{axis}
        [
        axis y line*=right,
        axis x line = none,
        yticklabel pos=right,
        ymin = 0.00,
        ymax = 1600,
        xmin = 4,
        xmax = 7.5,
        ylabel={period -- $T$}, 
        xlabel={$\alpha$},
        tick scale binop=\times,
        width=0.42\textwidth,
        height=0.35\textwidth,
        max space between ticks=25,
        ytick={0,400,800,1200,1600},
        ]
        \addplot+[no marks, thick, color=red!50!white] table[x
        expr=\thisrowno{0}, y expr=\thisrowno{1}, col sep=space] {Figures/alpha-per-limit.dat};
        \node at (axis cs:6.1, 360) {{\footnotesize \color{red!50!white} period $T$:  right axis $\rightarrow$}};
      \end{axis}
    \end{tikzpicture}
  }
  \subfigure[]{
    \begin{tikzpicture} 
      \begin{axis}
        [ 
        ymin = 0.00,
        ymax = 0.80,
        xmin = 0.4,
        xmax = 1.4,
        ylabel={amplitude -- $r_o^{\text{max}}$}, 
        xlabel={$\zeta$},
        tick scale binop=\times,
        width=0.42\textwidth,
        height=0.35\textwidth,
        grid=major,
        max space between ticks=25,
        ]
        \addplot+[no marks, thick, color=green!80!black] table[x
        expr=\thisrowno{0}, y expr=\thisrowno{1}, col sep=space] {Figures/zeta-amp-limit.dat};
        \node at (axis cs:1.38, 0.6) {\begin{minipage}{0.25\textwidth}{\footnotesize \color{green!80!black} amplitude $r_o^{\text{max}}$\\  left axis $\leftarrow$}\end{minipage}};
      \end{axis}
      \begin{axis}
        [
        axis y line*=right,
        axis x line = none,
        yticklabel pos=right,
        ymin = 0.00,
        ymax = 800,
        xmin = 0.4,
        xmax = 1.4,
        ylabel={period -- $T$}, 
        xlabel={$\zeta$},
        tick scale binop=\times,
        width=0.42\textwidth,
        height=0.35\textwidth,
        max space between ticks=25,
        ]
        \addplot+[no marks, thick, color=green!80!black!50!white] table[x
        expr=\thisrowno{0}, y expr=\thisrowno{1}, col sep=space] {Figures/zeta-per-limit.dat};
        \node at (axis cs:0.98, 300) {{\footnotesize \color{green!80!black!50!white} period $T$:  right axis $\rightarrow$}};
      \end{axis}
    \end{tikzpicture}
  }
  \subfigure[]{
    \begin{tikzpicture} 
      \begin{axis}
        [ 
        ymin = 0.00,
        ymax = 0.80,
        xmin = 0.0065,
        xmax = 0.0115,
        ylabel={amplitude -- $r_o^{\text{max}}$}, 
        xlabel={$d_T$},
        tick scale binop=\times,
        width=0.42\textwidth,
        height=0.35\textwidth,
        grid=major,
        max space between ticks=25,
        ]
        \addplot+[no marks, thick, color=blue] table[x
        expr=\thisrowno{0}, y expr=\thisrowno{1}, col sep=space] {Figures/dT-amp-limit.dat};
        \node at (axis cs:0.0092, 0.6) {\begin{minipage}{0.25\textwidth}{\footnotesize \color{blue} amplitude $r_o^{\text{max}}$\\  left axis $\leftarrow$}\end{minipage}};

      \end{axis}
      \begin{axis}
        [
        axis y line*=right,
        axis x line = none,
        yticklabel pos=right,
        ymin = 0.00,
        ymax = 800,
        xmin = 0.007,
        xmax = 0.0115,
        ylabel={period -- $T$}, 
        xlabel={$d_T$},
        tick scale binop=\times,
        width=0.42\textwidth,
        height=0.35\textwidth,
        max space between ticks=25,
        ]
        \addplot+[no marks, thick, color=blue!50!white] table[x
        expr=\thisrowno{0}, y expr=\thisrowno{1}, col sep=space] {Figures/dT-per-limit.dat};
        \node at (axis cs:0.009, 220) {{\footnotesize \color{blue!50!white} period $T$:  right axis $\rightarrow$}};
      \end{axis}
    \end{tikzpicture}
  }
\caption{Amplitude of the oscillatory excursions and their periods for
  varying (a) $\alpha$, (b) $\zeta$, (c) $d_T$, for a case initiated with $\alpha = 5$, $\zeta = 0.7$, and $d_T = 0.01$.}\label{fig:limit-prms}
\end{center}
\end{figure}
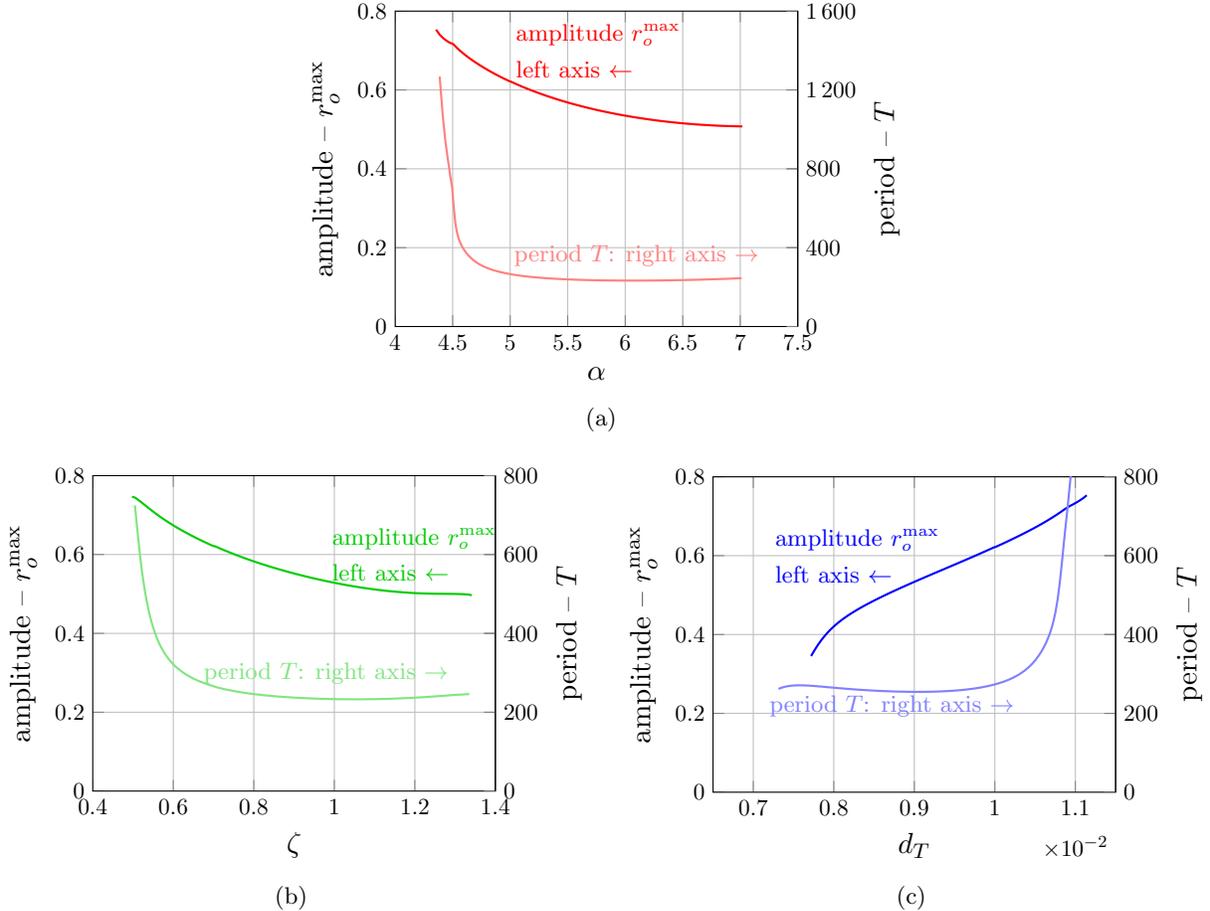

The amplitude of the oscillations and their period both depend on the parameters, with some curious observations about these dependencies seen in figure~\ref{fig:limit-prms}.   Starting from the same basic limit-cycle case we have been considering, with $\alpha = 5$, $\zeta = 0.7$, and $d_T= 0.01$, each parameter is varied between the bounds for which limit-cycle behavior persists.  Of course, only certain circumstances, typically by chance after a chaotic evolution, end up in the limit-cycle rather than the fixed point case, and arriving there can take significant simulation time.  Hence, studying the dependence of the limit-cycle behavior by slowly adjusting parameters, starting from an established limit-cycle, is more straightforward than initializing new simulations for each set of parameters.  Hundreds of oscillations occur across the full range for all of these parameter sweeps.

Figure~\ref{fig:limit-prms} (a) shows that the stability for the cyclic behavior fails when $\alpha$ is slowly increased above $\alpha \approx 7$, leaving a chaotic flow.  Decreasing the $\alpha$ activity strength below $\alpha \approx 4.3$ stabilizes the entire flow, leaving the object near the largest radius of the final cycle.

In figure~\ref{fig:limit-prms} (b), increasing $\zeta$ beyond about $\zeta = 1.3$ causes a failure of the limit-cycle behavior.  The subsequent chaotic flow persists until it arrives, after a stochastic process per the examples of section~\ref{s:initialtraj}, in a defect-free $m_\circ=+1$ configuration and fixed-point behavior, which persists for arbitrary large $\zeta$.  Decreasing $\zeta$ causes the oscillations to slow and increase in amplitude, until the limit-cycle fails to persist below $\zeta \approx 0.5$.  Decreasing $\zeta$ further causes the flow to cease altogether for $\zeta \lesssim 0.2$, though it remains in a nematically aligned state with $\lambda_1 < 0.1$, with the object now fixed in place.  Finally, the suspension becomes isotropic ($\bD \to \bI/2$) for $\zeta \lesssim 0.08$, consistent with the $2 \zeta/d_R=8$ bifurcation \cite{Gao:2017}.

\begin{figure}
  \begin{center}
    \subfigure[]{
    \begin{tikzpicture}
      \node[anchor=south west, inner sep=0] (image) at (0,0)
      {\includegraphics[width=0.4\textwidth]{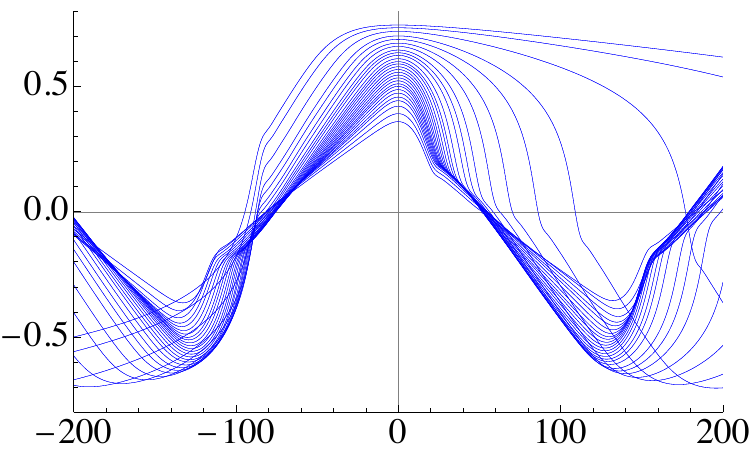}};
      \begin{scope}[x={(image.south east)},y={(image.north west)}]    
        \node at (0.5,-0.07) {$t-t_{\text{peak}}$}; 
        \node[rotate=90] at (-0.025,0.5) {$r_s$};
        \node[red] at (0.53,0.4) {increasing $d_T$};
        \draw [-stealth, red, thick](0.57,0.46) -- (0.65,0.8); 
      \end{scope}
    \end{tikzpicture}
    }
    \subfigure[]{ %
    \begin{tikzpicture}
      \node[anchor=south west, inner sep=0] (image) at (0,0)
      {\includegraphics[width=0.4\textwidth]{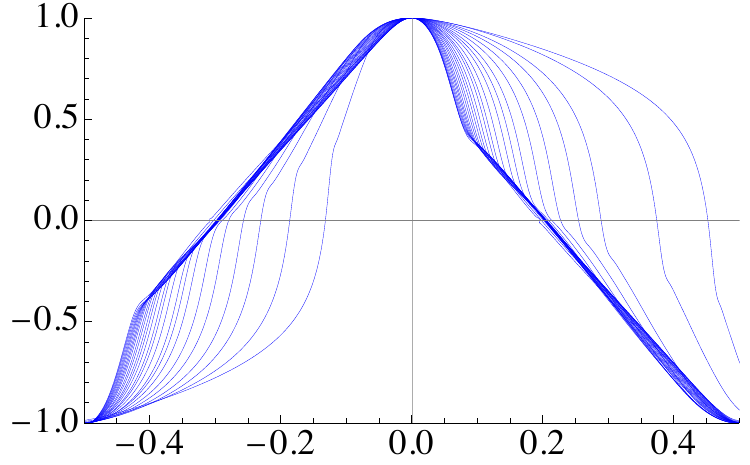}};
      \begin{scope}[x={(image.south east)},y={(image.north west)}]    
        \node at (0.5,-0.07) {$(t-t_{\text{peak}})/T_{\text{period}}$}; 
        \node[rotate=90] at (-0.025,0.5) {$r_s/r_{\text{max}}$};
        \node[red] at (0.6,0.4) {increasing $d_T$};
        \draw [-stealth, red, thick](0.2,0.3) -- (0.5,0.5);
      \end{scope}
    \end{tikzpicture}
    }
  \end{center}
  \caption{Histories of signed object radius ($r_s = r_o \sgn x$) for values of $d_T$ ranging from
    $0.00775$ to $0.01120$ in intervales of $\Delta d_T = 0.000125$:  (a) time-shifted data and (b) the same normalized by the amplitude and period.  In all cases, the large $d_T$ values correspond to peaks that extend more broadly to the right.}\label{fig:dTtraj}
\end{figure}

The dependence of amplitude on $d_T$ is particularly curious, with a factor of two change in amplitude across the relatively narrow range of $d_T$ from 0.0077 to 0.0110.  That the amplitude increases for increased $d_T$ diffusion might also be particularly unexpected.  The trajectories for this full $d_T$ range are shown in figure~\ref{fig:dTtraj} (a), with displacement histories for different $d_T$ values aligned at their peaks.  When also normalized by the peak value and the period in figure~\ref{fig:dTtraj} (b), there is a remarkable collapse for parts of the trajectories, which are differentiated primarily by when the kink in the trajectory occurs.  (Note that these figures include the final motions before the oscillating pattern fails for $d_T \approx 0.011$, hence the imperfect final period.)  Figure~\ref{fig:hists} showed that the kink for a particular case, and the corresponding rapid slowing of the motion, occurs as $\lambda_1$ suddenly approach 0.5.  Increased $d_T$ diffusion suppresses fracture, allowing the faster flow to persist longer, decreasing the traverse time.  Increasing $d_T$ also increases the diffusion associated with the reversal, as discussed, but this is not its most consequential effect on the period because $d_T$ only changes a relatively small amount.

\section{More complex configurations}
\label{s:other}

The basic fixed-point and limit-cycle behaviors are distinct and easily characterized for the circular container, but they do not depend on this geometry.  
Indeed, in crafting this study, the limit cycle was first observed in a $3.8\times 3.8$ square container.  In this case, similar parameters ($\alpha = 4.8$, $\zeta = 0.625$) lead to limit-cycle oscillations on the diagonal of the square.  In general, this solution seems to take longer to arise since it depends on the nematic alignment being nearly diagonal within the square and the object being positioned near the corresponding symmetry axis, whereas only the second of these criteria is required in a circular container.  (An animation of this square-container case is available as supplementary material Movie~3.)  Similarly, arriving at a $m_\circ=+1$ condition in a square moves the object toward a nearby container wall.  If that is near the center one of the faces of the square, it nearly stops just off the wall as for the fixed-point in a circle, but it then slowly approaches a corner; the center of each face of the square is an unstable equilibrium point.  This behavior is observed, for example, for $\alpha = 3.5$ and $\zeta = 1$.   If the circle directly approaches the corner of the square, there is a less pronounced overshoot than observed for the circle (figures~\ref{fig:basictraj} a and b and figure~\ref{fig:fixedpoint-approach}), but it settles into a fixed point.  This was not yet observed in simulations, but is anticipated as a potential outcome since it was easy to construct this case with a corresponding initial condition.

\begin{figure}
\begin{center}
  \subfigure[]{
    \begin{tikzpicture}
      \begin{axis}
        [ 
        ymin = -0.05,
        ymax = 1.15,
        xmin = 0,
        xmax = 40000,
        ylabel={$r_o(t)$},
        xlabel={$t$},
        tick scale binop=\times,
        width=0.43\textwidth,
        height=0.35\textwidth,
        grid=major,
        max space between ticks=25, 
        ytick={0.0,0.2,0.4,0.6,0.8,0.9,1.0,1.1},
        yticklabels={0.0,0.2,0.4,0.6,0.8,0.9,,1.1},
        ]
        \addplot [samples=10, domain=0:40000,dashed] ( x, 0.9 );
        \addplot [samples=10, domain=0:40000,dashed] ( x, 1.1 );
        \addplot+[no marks, thick, line join=round , mesh, point meta=\thisrowno{0}, colormap/bluered,
        colormap={}{ %
           [1cm] color(0cm)=(red) color(1cm)=(green) color(2cm)=(blue)
        }] table[x 
        expr=\thisrowno{0}, y expr=\thisrowno{8}, col sep=space] {Figures/pp-ellipse-e100.dat};
      \end{axis}
    \end{tikzpicture}
  }
  \subfigure[]{
    \raisebox{-0.1in}{
    \begin{tikzpicture}
      \begin{axis}
        [ 
        ymin = -1,
        ymax = 1,
        xmin = -1.2,
        xmax = 1.2,
        xlabel={$x_o(t)$},
        ylabel={$y_o(t)$},
        tick scale binop=\times,
        width=0.44\textwidth,
        height=0.35\textwidth,
        xtick={-1,-0.5,0.0,0.5,1},
        ytick={-0.8,-0.4,0.0,0.4,0.8},
        ]
        \addplot+[no marks, thick, mesh, point meta=\thisrowno{0}, colormap/bluered,
        colormap={}{ %
           [1cm] color(0cm)=(red) color(1cm)=(green) color(2cm)=(blue)
        }] table[x 
        expr=\thisrowno{2}, y expr=\thisrowno{3}, col sep=space] {Figures/pp-ellipse-e100.dat};
        \addplot [samples=100, domain=0:2*pi,dotted] ( {1.1*cos(deg(x))}, {0.9*sin(deg(x))} );
      \end{axis}
    \end{tikzpicture}
    }
  }
\caption{Trajectories for $\alpha = 4.8$ and $\zeta = 1.0$ in an elliptical container with 2.2 and 1.8 semi-major and semi-minor axes:  (a) $r_o(t)$ radius history with major- and minor-axis contact radii shown with dashed lines, and (b) the trajectory, with the dotted curve indicating the would-be contact.  The same color pattern tracks evolution in
  time in both frames.  }\label{fig:ellipse}
\end{center}
\end{figure}
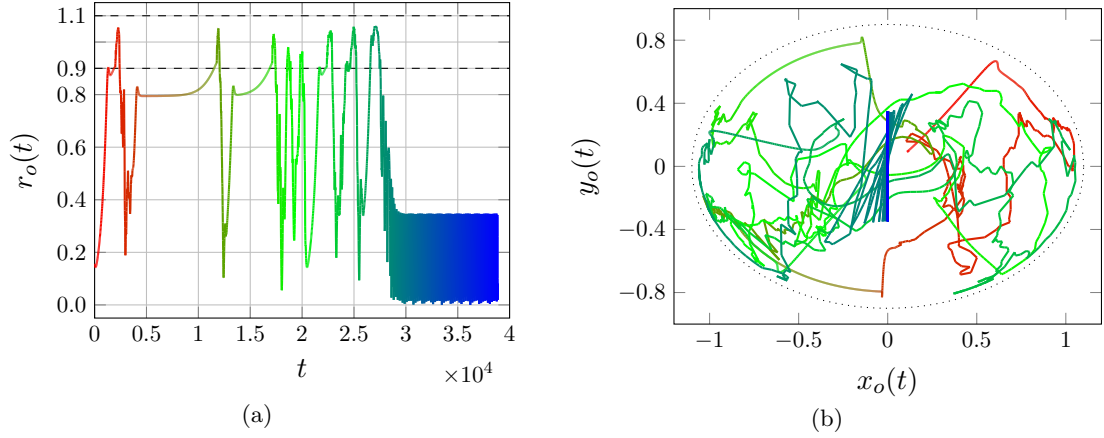

Still richer phenomenology is possible.  Figure~\ref{fig:ellipse} (and supplemental Movie~4) shows the same object in an elliptical container.  In this case, the object often moves chaotically, but as in the case of the circle, it can arrive in a defect-free $m_\circ=+1$ radial state, which brings it toward the nearest wall.   However, rather than becoming fixed, the solution enters a slow evolution phase.  The local asymmetry of the ellipse-shaped container causes slow motion toward higher container-wall curvature.  However, before reaching the point of maximum wall curvature, the overall suspension destabilizes, lifting the object off the wall.  This repeats multiple times, with considerable time spent in near-fixed-point configurations.  Remarkably, after 5 clear such cycles, the object then ends up in a $m_\circ=0$ condition without defects, and starts to oscillate.  It achieves a limit cycle after the axis of the oscillations slowly rotates aligns with the minor axis of the container.  Decreasing the activity strength from $\alpha = 4.8$ to 4.5 preserves stability of the fixed-point-like case even as it approaches the maximum radius, where it stops, as in the circle cases, but now at a point of minimum curvature.  Hence, this active fluid can place the object in a fixed location in a container regardless of its starting point even with persistent flow.  Large enough $\zeta$ can seemingly preclude the limit-cycle behavior of figure~\ref{fig:limit-prms} (b) to guarantee this.

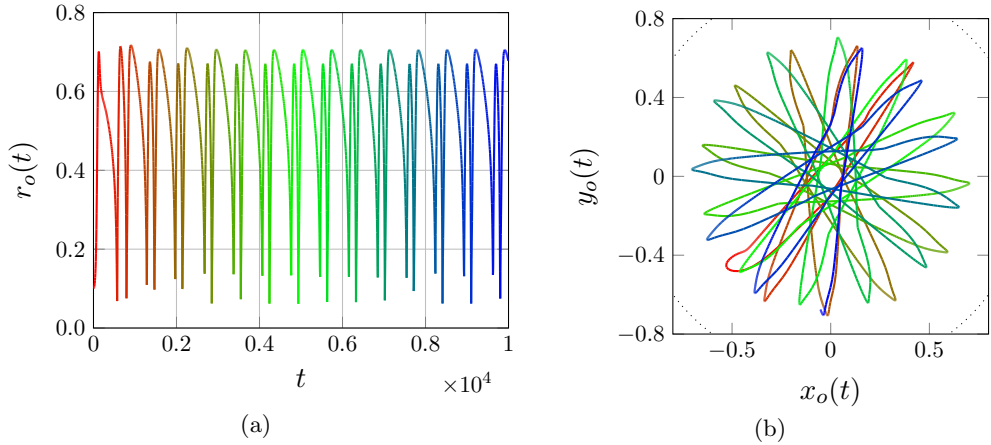
\begin{figure}
\begin{center}
  \subfigure[]{
    \begin{tikzpicture}
      \begin{axis}
        [ 
        ymin = -0.0,
        ymax = 0.8,
        xmin = 0,
        xmax = 10000,
        ylabel={$r_o(t)$},
        xlabel={$t$},
        tick scale binop=\times,
        width=0.43\textwidth,
        height=0.35\textwidth,
        grid=major,
        max space between ticks=25, 
        ytick={0.0,0.2,0.4,0.6,0.8},
        yticklabels={0.0,0.2,0.4,0.6,0.8},
        ]
        \addplot [samples=10, domain=0:40000,dashed] ( x, 0.9 );
        \addplot [samples=10, domain=0:40000,dashed] ( x, 1.1 );
        \addplot+[no marks, thick, line join=round, , mesh, point meta=\thisrowno{0}, colormap/bluered,
        colormap={}{ %
           [1cm] color(0cm)=(red) color(1cm)=(green) color(2cm)=(blue)
        }] table[x 
        expr=\thisrowno{0}, y expr=\thisrowno{8}, col sep=space] {Figures/pp-spirograph.dat};
      \end{axis}
    \end{tikzpicture}
  }
  \subfigure[]{
    \raisebox{-0.1in}{
    \begin{tikzpicture}
      \begin{axis}
        [ 
        ymin = -0.8,
        ymax = 0.8,
        xmin = -0.8,
        xmax = 0.8,
        xlabel={$x_o(t)$},
        ylabel={$y_o(t)$},
        tick scale binop=\times,
        width=0.35\textwidth,
        height=0.35\textwidth,
        xtick={-1,-0.5,0.0,0.5,1},
        ytick={-0.8,-0.4,0.0,0.4,0.8},
        ]
        \addplot+[no marks, thick, mesh, point meta=\thisrowno{0}, colormap/bluered,
        colormap={}{ %
           [1cm] color(0cm)=(red) color(1cm)=(green) color(2cm)=(blue)
        }] table[x 
        expr=\thisrowno{2}, y expr=\thisrowno{3}, col sep=space] {Figures/pp-spirograph.dat};
        \addplot [samples=100, domain=0:2*pi,dotted] ( {1.*cos(deg(x))}, {1.0*sin(deg(x))} );
      \end{axis}
    \end{tikzpicture}
    }
  }
\caption{Trajectories for $\alpha = 4.8$, $\zeta = 0.625$ and $d_T = 0.104$ in a circular container:  (a) radius history, and (b) spatial trajectories with the dotted curve indicating would-be contact.  The same color pattern tracks evolution in
  time in both frames.  }\label{fig:spirograph}
\end{center}
\end{figure}

For some narrow ranges of parameters, the circular container itself can also show modestly more complex behavior, such as in figure~\ref{fig:spirograph}, which is an oscillatory solution with an additional precession and missing of the $r=0$ point.  A similar behavior also signals the first breakdown of the center-passing limit cycle with decreasing $d_T$, such as in figure~\ref{fig:limit-prms} (c).  Decreasing $d_T$ to $d_T \lesssim 0.0078$ causes the oscillations to no longer pass through $r=0$, but to persist with an only modest change in their $r$ extent.  Note, however, that this solution also has a period doubling, in that every other pass reaches the maximum $r_o$, which suggest a bifurcation.  Given that these solutions seem rare relative to the basic cases, their importance is probably most relevant to dynamical system analysis, which is deferred.

\section{Additional Discussion}
\label{s:conclusions}

The behaviors found and analyzed show rich phenomenology for an object within a confined active fluid.  That they are robust to parameter changes and seem to always yield one of the outcomes supports their potential importance, as does the appearance of similar behaviors in non-circular containers.  The behaviors we see depend on long-range nematic ordering, into either the energy minimum $\theta$ constant state with $m_\circ =0$ or the non-equilbrium local energy minimum radial $\theta = \phi$ state with $m_\circ = +1$.  Hence, both of the behaviors depend on strong (fast $\tau_\zeta$) nematic ordering.  However, they also involve the interplay of alignment effects with small-scale (fast $\tau_T^\delta$) diffusion, and the interplay of object advection (slow $\tau_{u_\circ}^\ell$) and large-scale diffusion (slow $\tau_T^\ell$).  The time scale of the active flow ($\tau_{u_f}^\ell$) that results is intermediate to these, which seems to underpin the richness of the observed behaviors.  Significant flow is necessary for deforming the nematic field in the limit-cycle behavior, as for the chaotic wanderings before arriving at either of the observed outcomes.  
Of course, the actual biological or biology-inspired fluids that motivate the model here are far more complex, both in their local structure and their inhomogeneity.  How their character might be adjusted, locally or in time, to achieve structures or accomplish tasks is the overarching goal of these investigations.  The observations here are viewed as building blocks in this.  Any experiment will also include three-dimensional effect, even in cases where the flow is nominally two-dimensional.  Still, we can anticipate many of the features shown here might apply.  

The two primary solutions were not discovered by any systematic investigations.  Rather, they were in truth stumbled upon through observation.  Neither was pre-imagined by this author.  Additional behaviors might be waiting to be discovered for this or related geometries, though it is hard to imagine what their character would be, and long-time simulations did not discover any for the ranges of parameters studied.  A systematic search, either with fine scanning of physical parameters, initial conditions, or container or object shapes might turn up others for simple geometries and likely would for more complex geometries.  Using the adjoint governing equations to provide sensitivity fields \cite{Freund:2018} might speed discovery.  Additional boundary conditions, particularly wall-constrained nematic ordering, is a potentially important direction of further investigation, and enable mediated processes and additional outcomes.  It is interesting that the primary behaviors analyzed arise out of chaotic disordered states, which suggests that dynamical systems descriptions might provide more general insights into their behavior and design.

\subsection*{Acknowledgments}

The author is grateful for the thoughtful insights and inciscive questions of Randy Ewoldt in discussing this work and commenting on a draft. 

\bibliographystyle{my-elsarticle-num}%
\bibliography{activesus.bib}

\end{document}